\documentclass[11pt]{article}

\usepackage{amsmath} 
\usepackage{amssymb}
\usepackage{bm} 

\usepackage{upgreek}
\usepackage{verbatim} 

\usepackage{jheppub}

\hypersetup{colorlinks=true,citecolor=blue,linkcolor=black,urlcolor=blue}

\numberwithin{equation}{section}

\newcommand{\de}{\partial}
\newcommand{\lr}[1]{\left(#1\right)}
\newcommand{\pder}[2]{\frac{\partial #1}{\partial #2}}

\newcommand{\xiv}{{\bm \xi}}

\def\Re{{\rm Re}}
\def\Im{{\rm Im}}
\def\k{{\bf k}}
\def\ce{\varepsilon}

\def\A{\mathtt{A}}

\def\mmu{\upmu}
\def\muphi{{\mu_{\scriptscriptstyle \Phi}}}
\def\rhophi{\rho_{\scriptscriptstyle \Phi}}
\def\coeff#1#2{{\textstyle {\frac {#1}{#2}}}}
\def\pperp{{\scriptscriptstyle \perp}}
\def\pparallel{{\scriptscriptstyle \parallel}}

\title{Causality in dissipative relativistic magnetohydrodynamics}
\author{Raphael E. Hoult, Pavel Kovtun}
\affiliation{
{\it\small Department of Physics \& Astronomy, University of Victoria}\\
{\it\small PO Box 1700 STN CSC, Victoria, BC, V8W 2Y2, Canada}\\
}
\emailAdd{rhoult@uvic.ca, pkovtun@uvic.ca}
\abstract{We explore the relationship between linear and non-linear causality in theories of dissipative relativistic fluid dynamics. While for some fluid-dynamical theories, a linearized causality analysis can be used to determine whether the full non-linear theory is causal, for others it can not. As an illustration, we study relativistic viscous magnetohydrodynamics supplemented by a neutral-particle current, with resistive corrections to the conservation of magnetic flux. The dissipative theory has 10 transport coefficients, including anisotropic viscosities, electric resistivities, and neutral-particle conductivities. We show how causality properties of this magnetohydrodynamic theory, in the most general fluid frame, may be understood from the linearized analysis.}

\begin{document}
\maketitle

\section{Introduction}
Hydrodynamics is amongst the oldest fields in physics; nevertheless, its foundational questions still form an active area of research. Here we are interested in relativistic dissipative hydrodynamics, which is a classical macroscopic description of fluids consistent with both special and general relativity~\cite{Rezzolla-Zanotti, Romatschke:2017ejr}. Relativistic effects may be important for matter at very high temperatures and very high densities; hence, relativistic hydrodynamics has found applications in describing the quark-gluon plasma created in heavy-ion collisions~\cite{Romatschke:2017ejr}, accretion disks around black holes~\cite{Abramowicz:2011xu}, and binary neutron star mergers~\cite{Faber:2012rw}.

It is by now well understood that including the effects of macroscopic dissipation (viscosity, heat conduction) into the relativistic hydrodynamic framework is plagued with uncertainties. The first formulations of dissipative relativistic equations were introduced by Eckart~\cite{PhysRev.58.919} in 1940, and by Landau and Lifshitz in the 1953 edition of their book ``Mechanics of continuous media'', later reissued as ``Hydrodynamics''~\cite{LL6}. While the hydrodynamic equations in both of these theories are relativistically covariant, both theories predict that uniformly moving equilibrium states are unstable~\cite{Hiscock:1985zz}, and moreover both theories predict super-luminal propagation~\cite{Hiscock:1987zz}. This can be traced to the fact that for a set of partial differential equations, relativistic covariance does not necessarily imply hyperbolicity. 

The above problems were addressed in the M\"uller-Israel-Stewart (MIS) theory~\cite{Muller:1967zza,  Israel:1976tn, Israel-Stewart}, which adds extra non-hydro\-dyna\-mic fields (fluxes) to the naive hydrodynamic theory; the independent relaxational dynamics of these extra fields is to be chosen so that the resulting equations do not lead to unphysical predictions. One can devise many variants of MIS-type theories because there many ways to introduce relaxational dynamics for the extra non-hydrodynamic fields. One can derive MIS-type equations motivated by the entropy current~\cite{Hiscock:1983zz}, by the form of the differential equations~\cite{Geroch:1990bw}, by general symmetry considerations~\cite{Baier:2007ix}, by kinetic theory~\cite{Denicol:2012cn}, or by the hydrostatic partition function~\cite{Jain:2023obu}.
Alternatively, the problems of the classic theories of \cite{PhysRev.58.919, LL6} may be rectified by the Bemfica-Disconzi-Noronha-Kovtun (BDNK) approach~\cite{Bemfica:2017wps, Kovtun:2019hdm, Bemfica:2019knx, Hoult:2020eho, Bemfica:2020zjp}, which does not introduce extra non-hydrodynamic fields; rather, one works with hydrodynamic fields (temperature, velocity, etc) whose out-of-equilibrium definitions are chosen to ensure hyperbolicity. In the BDNK approach, the only equations of hydrodynamics are the conservation laws; in the MIS approach, the conservation laws must be supplemented by relaxational equations for the extra non-hydrodynamic degrees of freedom. 

Our focus in this paper will be on relativistic dissipative magnetohydrodynamics (MHD), a classical theory of fluids made of electrically charged constituents. In MHD, electromagnetic fields themselves become hydrodynamic variables (in addition to the fluid velocity, etc)~\cite{Jackson2E}. The conservation equations of standard hydrodynamics thus need to be supplemented by the dynamical equations which determine the classical evolution of the electromagnetic fields. 
While perfect-fluid MHD can be made consistent with relativity \cite{Anile}, a naive introduction of dissipation into the relativistic MHD equations leads to exactly the same problems as one finds in standard relativistic hydrodynamics. 
In order to make the equations of relativistic MHD (with all first-order dissipative transport coefficients taken into account) causal, one needs to modify the naive MHD equations, either in an MIS-like way \cite{Cordeiro:2023ljz}, or in a BDNK-like way \cite{Armas:2022wvb}, or in some other way which restores causality.

It is important to mention that there are at least two philosophically different approaches to writing down the equations of relativistic dissipative MHD, even before one starts worrying about causality. In one approach, MHD is viewed as normal-fluid hydrodynamics coupled to Maxwell's equations in matter. Dissipation is then introduced through constitutive relations for the fluxes of energy, momentum, and particles, such as the Ohm's law $J^i_{\rm el.} = \sigma E^i + \dots$ for the electric current density $J^i_{\rm el.}$, where $E^i$ is the electric field. In this approach, electrical conductivities, such as $\sigma$, are the natural transport coefficients, in addition to the viscosities~\cite{Hernandez:2017mch}. Alternatively, one can view MHD as normal-fluid conservation equations coupled to the conservation equation of the magnetic flux density, whose corresponding conserved current (the flux of the magnetic flux density)
is an antisymmetric tensor. Dissipation is introduced through constitutive relations for the fluxes, now including the magnetic flux density $J^{ik}$, such as $J^{ik} = r (\partial^k B^i - \partial^i B^k) + \dots$, where $B^i$ is the magnetic field. In this approach, electrical resistivities, such as $r$, are the natural transport coefficients, in addition to the viscosities~\cite{Grozdanov:2016tdf}. For linear perturbations, the spectra of MHD waves are the same in the two approaches, and the transport coefficients can be related to each other~\cite{Hernandez:2017mch}. However, the non-linear evolution equations do differ in the two formulations. 

In this paper, we will study relativistic MHD formulated using the conservation equation for the magnetic flux with resistive corrections included; we will refer to this formulation of the theory as dMHD (``d'' for ``dual''), to differentiate from MHD formulated using Maxwell's equations in matter.  We will explore how the BDNK approach can be used to make the dissipative dMHD equations stable and causal. This question was addressed recently in reference~\cite{Armas:2022wvb} on the basis of dispersion relations for linearized near-equilibrium perturbations. It is, however, important to emphasize that the linearized causality analysis is in general not sufficient to conclude that the theory is causal: one can easily write down equations where the linearized near-equilibrium dynamics is causality-preserving, while the full non-linear dynamics is causality-violating. We will show that if the BDNK recipe applied to dMHD ensures that the resulting theory is causal linearly, it will also be causal non-linearly; this is a consequence of a more general result which applies to many BDNK-type theories. We include the neutral-particle current%
\footnote{
  In relativistic theories (unlike in Galilean theories), there is no law of mass conservation. The relevant conserved quantities in relativistic theories are charges which generate global symmetries (such as the $U(1)$ baryon number symmetry), together with the corresponding conserved currents. In this paper, we consider magnetohydrodynamics supplemented by such a conserved $U(1)$ global symmetry current $J^\mu$, which we will call the ``current of electrically neutral particles'', or ``neutral-particle current'' for short. A particle-like interpretation of the constituents of $J^\mu$ is not necessary in hydrodynamics; when the constituents of $J^\mu$ are actually particles of mass $m$, one can form a current $m J^\mu$ which may be (somewhat imprecisely) called a ``mass current''. 
}
in the BDNK-dMHD theory, and outline the derivation of the causality conditions. More generally, we will discuss the types of classical theories for which the linearized causality analysis is sufficient to make the non-linear theories causal. 

The outline of the paper is as follows. In section~\ref{sec:caus_and_eq}, we discuss the equivalence between causality of linearized equations vs causality of the actual non-linear equations; we discuss conditions under which the dispersion relations of the linearized theory correctly reflect the causality of the non-linear theory. In section~\ref{sec:intro-MHD}, we introduce the magnetohydrodynamic theory formulated using the conservation equation for the magnetic flux density. Starting with perfect-fluid MHD, we proceed to perfect-fluid dMHD, and then to dissipative dMHD.  In section~\ref{sec:equiv_dMHD}, we discuss causality restoration in non-linear dissipative dMHD according to the BDNK recipe; following the criteria of section~\ref{sec:caus_and_eq}, we find that the linearized causality analysis in dissipative dMHD is indeed sufficient. 
In section~\ref{sec:conclusion}, we summarize our results and look to future applications.

\section{Linear vs non-linear causality}
\label{sec:caus_and_eq}
To begin, we note that dissipative first-order hydrodynamic theories discussed in references~\cite{Kovtun:2019hdm, Bemfica:2019knx} have the following property:
\begin{align}
\label{equivalence}
  \parbox{0.8\textwidth}{\em If the equations of motion linearized about a constant solution in flat space are causal, then the non-linear theory is also causal.}
\end{align}
In order to understand why this is the case, consider a system of second-order partial differential equations of the form
\begin{equation}
\label{eq:quasilin_gen}
    {\cal A}^{\mu\nu}_{AB}[U, \de U] \de_\mu \de_\nu U^B + {\cal B}^{\mu}_{AB}[U, \de U] \de_\mu U^B + {\cal C}_A[U] = 0\,,
\end{equation}
where capital Latin indices run from $1$ to $n$ (the number of unknown functions), Greek indices label the spacetime coordinates, and $U^B$ is a vector of the dynamical degrees of freedom, possibly including the dynamical metric. The first term in equation~\eqref{eq:quasilin_gen} is referred to as the ``principal part". 
The dissipative hydrodynamic theories of references~\cite{Kovtun:2019hdm, Bemfica:2019knx} have the general form of equation~\eqref{eq:quasilin_gen}, with two additional properties: ${\cal C} = 0$, and ${\cal A}^{\mu\nu}_{AB}$ does not depend on $\partial U$. These are due to the fact that the only hydrodynamic equations are conservation laws, and that the constitutive relations are written in the derivative expansion, to first order in derivatives.%
\footnote{
  There are examples of BDNK-type theories which don't have these two properties, such as the $\ce_3 = \pi_3 = \nu_3 = \theta_3 = \gamma_3 = 0$ frame described in \cite{Abboud:2023hos}. In this hydrodynamic frame, the system becomes mixed-order in derivatives, and derivative-dependence enters the principal part. Since the system is mixed order, it is not of the form \eqref{eq:quasilin_gen}; we will not be considering such systems in this paper.
  } 
This may be contrasted with MIS-type theories~\cite{Muller:1967zza,  Israel:1976tn, Israel-Stewart}, where ${\cal C}[U]$ is always non-zero. 

Using the standard terminology, we will often refer to a given system of differential equations as a ``theory''. The causality of the theory~\eqref{eq:quasilin_gen} is governed by the characteristics of the system, which are families of hypersurfaces in spacetime along which initial conditions cannot be consistently provided. Characteristics are important because they are the only hypersurfaces across which the solution can experience a discontinuity. As such, they serve as wavefronts for the theory, and determine the speed of signal propagation~\cite{Courant-Hilbert}.
The characteristics are found by solving the characteristic equation associated with \eqref{eq:quasilin_gen}, 
\begin{equation}
\label{eq:char_eq}
    Q \equiv \det\lr{{\cal A}^{\mu\nu}[U,\de U]\xi_\mu \xi_\nu} = 0 \,,
\end{equation}
where $\xi_\mu = (\xi_0, \xiv)$ are co-vectors which are normal to the characteristics. In a system with a finite number of partial differential equations such as \eqref{eq:quasilin_gen}, the characteristic equation $Q=0$ is a finite-order polynomial equation in the components of $\xi_\mu$ (we assume that $Q$ is not identically zero). In order for the system to be causal (meaning the characteristics are inside or on the lightcone), the following three conditions must be satisfied for real $\xiv$~\cite{Hoult:2023clg}:
\begin{equation}
\label{eq:non-linear_caus_conds_xi}
    \frac{|\Re\lr{\xi_0(\xiv)}|}{|\xiv|} \leq 1, \qquad \Im(\xi_0(\xiv)) = 0, \qquad {\cal O}_{\xi_0}(Q(\xi_\mu)) = {\cal O}_{|\xiv|}(Q(\xi_0 = a |\xiv|, \xiv_j = s_j |\xiv|)) \,.
\end{equation}
Here ${\cal O}_\chi(Q)$ denotes the order of the polynomial $Q$ in a variable $\chi$, $a\neq 0$, and $s_j$ is a constant unit vector.
The first condition is due to the demand that the characteristics lie inside or on the lightcone (and therefore their normals point outside or on the lightcone), while the second and third demands are to ensure (at least weak) hyperbolicity. Taking $|\xiv| \neq 0$, then, the first two conditions of \eqref{eq:non-linear_caus_conds_xi} may be re-written in terms of $V_\mu \equiv \xi_\mu/ |\xiv|$ as
\begin{equation}
\label{eq:nonlin_caus_conds}
    |\Re(V_0)| \leq 1, \qquad \Im(V_0) = 0, \qquad {\cal O}_{\xi_0}(Q(\xi_\mu)) = {\cal O}_{|\xiv|}(Q(\xi_0 = a |\xiv|, \xiv_j = s_j |\xiv|)) .
\end{equation}
The statement \eqref{equivalence} may then be interpreted as the statement that if the conditions \eqref{eq:nonlin_caus_conds} are satisfied by the linearized perturbations, they are also satisfied in the non-linear theory. Let us now look at examples.

\subsection{Examples}
\label{sec:examples}
\paragraph{Example 1.}
\label{ex:ex1}
Consider the following differential equation for a real scalar field $\phi(x)$:
\begin{align}
\label{eq:ex-1}
  u^\mu u^\nu \de_\mu \de_\nu \phi - c^2(\phi) \Delta^{\mu\nu} \de_\mu \de_\nu \phi  = 0 \,,
\end{align}
where $u^\mu$ is a timelike unit vector, and $\Delta^{\mu\nu} = g^{\mu\nu} + u^\mu u^\nu$ projects onto the space orthogonal to $u$. In our later discussion of hydrodynamics, $u^\mu$ will be the ``fluid velocity'', endowed with its own dynamics; for now, we take $u^\mu$ as fixed. 
The characteristic equation is given by
\begin{equation}
\label{eq:Q11}
    Q = \lr{u{\cdot}\xi}^2 - c^2(\phi) \Delta^{\mu\nu}\xi_\mu \xi_\nu  = 0\,.
\end{equation}
This example is of course just the wave equation in a reference frame specified by $u^\mu$, and with propagation speed $c^2(\phi)$. The theory will be causal if the function $c^2(\phi)$ lies between $0$ and $1$, e.g.\ $c^2(\phi) = 1/(1+\phi^2)$. 

Now consider linearized perturbations about a constant solution $\phi_0$, i.e.\ $\phi(x) = \phi_0 + \delta \phi(x)$, 
\begin{align}
  u^\mu u^\nu \de_\mu \de_\nu \, \delta \phi - c^2(\phi_0) \Delta^{\mu\nu} \de_\mu \de_\nu \, \delta\phi = 0 \,.
\end{align}
The characteristic equation for the linearized problem is given by 
\begin{equation}
\label{eq:Q12}
    Q_{\rm lin} = \lr{u{\cdot}\xi}^2 - c^2(\phi_0) \Delta^{\mu\nu}\xi_\mu \xi_\nu  = 0\,.
\end{equation}
The characteristic equations \eqref{eq:Q11} and \eqref{eq:Q12} are algebraic equations for $\xi_\mu$, and the non-linear characteristic equation \eqref{eq:Q11} can be obtained from the linear characteristic equation \eqref{eq:Q12} by simply taking $\phi_0 \to \phi(x)$. Thus if the linearized perturbations are causal for all constant ``background'' values $\phi_0$, as judged by the $\phi_0$-dependent solutions of \eqref{eq:Q12}, then the non-linear theory \eqref{eq:ex-1} will also be causal. In other words, in this example the linearized analysis is sufficient to establish non-linear causality.

\paragraph{Example 2.}
As our next example, consider the case where the propagation speed depends on the gradients of the field,
\begin{align}
\label{eq:ex-2}
  u^\mu u^\nu \de_\mu \de_\nu \phi - c_s^2(\partial\phi) \Delta^{\mu\nu} \de_\mu \de_\nu \phi  = 0 \,,
\end{align}
with $c_s^2(\phi) = 1 + \Delta^{\mu\nu} \partial_\mu \phi \, \partial_\nu \phi$. The characteristic equation is
\begin{equation}
\label{eq:Q21}
    Q = \lr{u{\cdot}\xi}^2 - c_s^2(\partial\phi) \Delta^{\mu\nu}\xi_\mu \xi_\nu  = 0\,.
\end{equation}
When the equation is linearized about a constant solution $\phi_0$, i.e.\ $\phi(x) = \phi_0 + \delta \phi(x)$, the characteristic equation for the linearized theory is
\begin{equation}
\label{eq:Q22}
    Q_{\rm lin} = \lr{u{\cdot}\xi}^2 -  \Delta^{\mu\nu}\xi_\mu \xi_\nu  = g^{\mu\nu} \xi_\mu \xi_\nu = 0\,.
\end{equation}
The characteristics of the linearized theory happen to coincide with the actual light cone, hence the linearized theory is causal. On the other hand, the full characteristic equation \eqref{eq:Q21} implies that the actual theory breaks causality for any space-dependent $\phi(x)$. In other words, the linearized analysis can not be used to argue for the  causality of the theory.

\paragraph{Example 3.}
\label{ex:ex3}
Now consider the equation
\begin{align}
\label{eq:ex-3}
  u^\mu u^\nu \de_\mu \de_\nu \phi - c_s^2(\phi) \Delta^{\mu\nu} \de_\mu \de_\nu \phi +\phi = 0 \,,
\end{align}
with $c_s^2(\phi) = 1+\phi^2$. Unlike the previous example, there are no derivatives in the principal part; however, the equation now contains a term without derivatives.
The characteristic equation is
\begin{equation}
\label{eq:Q31}
    Q = \lr{u{\cdot}\xi}^2 - c_s^2(\phi) \Delta^{\mu\nu}\xi_\mu \xi_\nu  = 0\,.
\end{equation}
When equation~\eqref{eq:ex-3} is linearized about a constant solution $\phi_0$, the last term imposes that the only constant solution is $\phi_0=0$. Thus, linearization amounts to $\phi(x) = 0 + \delta\phi(x)$, and the resulting linear theory for $\delta\phi$ has the characteristic equation
\begin{equation}
\label{eq:Q32}
    Q_{\rm lin} = \lr{u{\cdot}\xi}^2 -  \Delta^{\mu\nu}\xi_\mu \xi_\nu  = g^{\mu\nu} \xi_\mu \xi_\nu = 0\,.
\end{equation}
Again, the characteristics of the linearized theory happen to coincide with the actual light cone, and the linearized theory is causal. On the other hand, the full characteristic equation \eqref{eq:Q31} implies that the non-linear theory breaks causality for any non-zero $\phi(x)$. Again, in this example the linearized analysis can not be used to argue for the non-linear causality of the theory.

\paragraph{Example 4.}
\label{ex:ex4}
Consider the equation
\begin{align}
\label{eq:ex-4}
  {\cal A}^{\mu\nu}(\lambda, \phi) \partial_\mu \partial_\nu \phi + {\cal C}(\phi) = 0 \,,
\end{align}
where the coefficient ${\cal A}$ of the principal part depends on parameters collectively denoted as $\lambda$. 
Demanding that the solutions to the characteristic equation
\begin{align}
\label{eq:Q41}
   Q = {\cal A}^{\mu\nu}(\lambda,\phi)\, \xi_\mu \xi_\nu = 0 
\end{align}
satisfy the causality conditions \eqref{eq:non-linear_caus_conds_xi} will in general place ``causality constraints'' on the parameters $\lambda$. For example, for ${\cal A}^{\mu\nu}(\lambda,\phi) = u^\mu u^\nu - (1+\phi^2)^\lambda \Delta^{\mu\nu}$, causality demands $\lambda < 0$; more generally, causality constraints will also involve $\phi$. As \nameref{ex:ex3} demonstrates, equations \eqref{eq:ex-4} with non-zero ${\cal C}(\phi)$ are not necessarily amenable to a linearized analysis which will determine the non-linear causality: for example, for a theory with ${\cal C}(\phi) = \phi$ and the above ${\cal A}^{\mu\nu}(\lambda,\phi)$, the linearized theory is always causal, while the non-linear theory is only causal for $\lambda<0$. The issue, however, is easy to fix: consider a ``partner'' equation
\begin{align}
\label{eq:ex-4a}
  {\cal A}^{\mu\nu}(\lambda, \chi) \partial_\mu \partial_\nu \chi  = 0 \,,
\end{align}
obtained from the original equation by simply removing the ${\cal C}(\phi)$ term. The solutions of the two equations \eqref{eq:ex-4} and \eqref{eq:ex-4a} are of course different. However, equation~\eqref{eq:ex-4a} always admits constant solutions $\chi = \chi_0$ with arbitrary $\chi_0$, and can be linearized, $\chi(x) = \chi_0 + \delta \chi(x)$. The characteristic equation for the linearized ``partner'' theory is 
\begin{align}
\label{eq:Q42}
   Q_{\rm lin} = {\cal A}^{\mu\nu}(\lambda,\chi_0)\, \xi_\mu \xi_\nu = 0 \,.
\end{align}
The characteristic equations \eqref{eq:Q41} and \eqref{eq:Q42} are algebraic equations for $\xi_\mu$, and the characteristic equation \eqref{eq:Q41} of the non-linear theory can be obtained from the characteristic equation \eqref{eq:Q42} of the linearized theory by simply taking $\chi_0 \to \phi(x)$. Thus the ($\phi$-dependent, in general) causality constraints in the original theory \eqref{eq:ex-4} can be obtained from the ($\chi_0$-dependent, in general) causality constraints in the linearized ``partner'' theory by simply taking $\chi_0 \to \phi(x)$. In other words, the linearized analysis in the ``partner'' theory is sufficient to establish the actual causality conditions in the original theory.  

The above argument also applies when the theory has $n$ fields $\phi^A(x)$. In this case, the constant solutions $\phi^A_0$ are in general not arbitrary, but are constrained by the relations ${\cal C}_A(\phi^1_0, \dots,  \phi^n_0) = 0$. The linearized theory, obtained by expanding about $\phi^A = \phi^A_0$, can not in general be used to argue for the causality of the full theory. Even though the causality conditions cannot be obtained from the linearized analysis of the actual theory, they can be obtained from the linearized analysis of the ``partner'' theory, obtained by removing ${\cal C}_A(\phi)$ from the equations.

\paragraph{Example 5.} 
In our previous examples, the failure of the linearized theory to reproduce the causality conditions of the full theory could be traced to either a) derivatives appearing in the principal part ${\cal A}^{\mu\nu}(\phi, \partial\phi)$, or b) constraints on the equilibrium values of the fields due to ${\cal C}(\phi)$. 
Now consider a system of equations
\begin{align}
\label{eq:ex-5}
  {\cal A}^{\mu\nu}_{AB}(\phi) \partial_\mu \partial_\nu \phi^B + {\cal B}^\mu_{AB}(\phi, \partial \phi) \partial_\mu \phi^B = 0 \,,
\end{align}
where ${\cal A}^{\mu\nu}$ does not depend of $\partial\phi$, and the non-derivative terms ${\cal C}_A(\phi)$ are absent (dissipative hydrodynamic theories of references~\cite{Kovtun:2019hdm, Bemfica:2019knx} are of this form). The linearized theory is obtained by expanding $\phi^A(x) = \phi^A_0 + \delta \phi^A(x)$, with arbitrary constant $\phi^A_0$. As dictated by the characteristic equation, the causality conditions obtained in the linearized theory give the causality conditions in the full non-linear theory upon taking $\phi^A_0 \to \phi^A(x)$. This, however, is only true if the linearized theory is obtained by expanding about the most general constant solution $\phi^A_0$, with all distinct non-zero components. If one derives the linearized theory by expanding $\phi^A(x) = \tilde \phi^A_0 + \delta \phi^A(x)$, where $\tilde\phi^A_0$ is constrained, the causality conditions obtained in the linearized theory will in general be incomplete, as illustrated in \nameref{ex:ex3} (which can be easily generalized to include more fields).

Can one get away with only linearizing about a subset $\tilde\phi^A_0$ of the constant solutions $\phi^A_0$, and still obtain the correct causality conditions of the full theory? The answer is yes, provided the transformation which generates the full $\phi^A_0$ from $\tilde\phi^A_0$ is a symmetry of the principal part. For a linear transformation $\phi^A \to G_{AB} \phi^B$, the principal part has a symmetry $G$ if
\begin{align}
\label{eq:A-symmetry}
  G {\cal A} (\phi) G^{-1} = {\cal A}(G\phi) \,.
\end{align}
Suppose now that one linearizes the equations about a non-zero $\tilde\phi^A_0$, such that there is a linear transformation $G$, with $\phi^A_0 = G_{AB} \tilde \phi^B_0$.
The characteristic equation in the theory linearized about an arbitrary non-zero $\phi^A_0$ then coincides with the characteristic equation linearized about $\tilde \phi^A_0$, namely
$\det [{\cal A}^{\mu\nu}(\phi_0) \xi_\mu \xi_\nu ] = \det [{\cal A}^{\mu\nu}(\tilde\phi_0) \xi_\mu \xi_\nu ]$. 

As an example, consider a theory of $n$ fields, and take $\tilde \phi^A_0 = (c, 0, \dots, 0)$, with arbitrary non-zero $\phi^1_0 = c$. Any vector $\phi^A_0$ can be obtained by applying an orthogonal transformation $G$ to $\tilde \phi^A_0$. Then, for any theory where ${\cal A}_{AB}$ can be written as ${\cal A}_{AB} = f(\bm{\phi}^2) \delta_{AB} + g(\bm{\phi}^2) \phi_A \phi_B$, with $\bm{\phi}^2 \equiv (\phi^1)^2 + \dots + (\phi^n)^2$, the property \eqref{eq:A-symmetry} will hold, thanks to the orthogonality of $G$. Hence, for theories with such ${\cal A}_{AB}$, the causality of the full non-linear theory can be deduced by studying the causality of the theory linearized about $\tilde\phi^A_0 = (c, 0, \dots, 0)$.

As another example of \eqref{eq:ex-5}, consider a theory of a timelike vector field $\beta^\mu$, 
\begin{align}
  {\cal A}^{\mu\nu}_{\rho\sigma}(\beta) \partial_\mu \partial_\nu \beta^\sigma + {\cal B}^\mu_{\rho\sigma}(\beta, \partial\beta) \partial_\mu \beta^\sigma = 0\,.
\end{align}
The characteristic equation takes the form $F(\beta^2, \beta{\cdot}\xi, \xi^2) = 0$, a polymonial equation in $\beta{\cdot}\xi$ and $\xi^2$, with $\beta^2$-dependent coefficients. The explicit form of the characteristic equation in first-order causal dissipative hydrodynamics can be found in reference~\cite{Hoult:2020eho}. There are no derivatives in the principal part, hence the linearized analysis can be used to establish the causality of the theory. The characteristic equation is Lorentz invariant, and the causality conditions can be obtained by working in a reference frame in which $\beta_0^\mu  = (c, 0, \dots, 0)$.

\paragraph{Example 6.}
Consider the theory \eqref{eq:quasilin_gen} for two fields $U^A=(\chi, \psi)$ with ${\cal C}_A=0$ and 
\begin{equation}
    {\cal A}^{\mu\nu}_{AB} = \begin{bmatrix}
        u^\mu u^\nu - c_1(\chi,\psi)\Delta^{\mu\nu} &  f(\partial\chi,  \partial \psi) u^\mu u^\nu\\
        0 & u^\mu u^\nu -  c_2(\chi,\psi)\Delta^{\mu\nu}
    \end{bmatrix} ,
\end{equation}
where $c_{1,2}(\chi, \psi)$ are real functions. Derivatives appear in the principal part; however, importantly, the off-diagonal term does not contribute to the characteristic equation,
\begin{equation}
\label{works_ex_derivs}
    Q =  \lr{\lr{u{\cdot}\xi}^2 - c_1(\chi,\psi) \Delta^{\mu\nu}\xi_\mu \xi_\nu}\lr{\lr{u{\cdot}\xi}^2 - c_2(\chi,\psi) \Delta^{\mu\nu}\xi_\mu \xi_\nu} = 0 \,.
\end{equation}
The characteristic equation in the theory linearized about the constant solution $(\chi_0, \psi_0)$ is 
\begin{equation}
\label{works_ex_derivs_lin}
    Q_{\rm lin} =  \lr{\lr{u{\cdot}\xi}^2 - c_1(\chi_0,\psi_0) \Delta^{\mu\nu}\xi_\mu \xi_\nu}\lr{\lr{u{\cdot}\xi}^2 - c_2(\chi_0,\psi_0) \Delta^{\mu\nu}\xi_\mu \xi_\nu} = 0 \,.
\end{equation}
Thus, the causality conditions derived in the linearized theory, expressed in terms of $\chi_0$, $\psi_0$, give the actual causality conditions upon taking $\chi_0 \to \chi(x)$, $\psi_0 \to \psi(x)$, in spite of the derivatives appearing in the principal part of the equation. It is the characteristic equation, not the principal part per se, which determines the causality of the theory. 

\paragraph{Example 7.}
\label{ex:ex7}
Returning to the wave equation in \nameref{ex:ex1}, let us define auxiliary fields $\chi_\mu \equiv \de_\mu \phi$. The wave equation \eqref{eq:ex-1} may then be re-written as a first-order system,
\begin{subequations}
\label{eq:we-phichi}
    \begin{align}
\label{eq:we-phichi-1}
        u^\mu u^\nu \de_\mu \chi_\nu - c^2(\phi) \Delta^{\mu\nu}\de_\mu \chi_\nu &= 0 \,,\\
\label{eq:we-phichi-2}
        \de_\mu \chi_\nu - \de_\nu \chi_\mu &=0 \,,\\
        \de_\mu \phi - \chi_\mu &= 0\,.
\label{eq:we-phichi-3}
    \end{align}
\end{subequations}
In $D$ spacetime dimensions, there are then $D{+}1$ functions $(\chi_\mu, \phi)$ to be determined. Ignoring equation~\eqref{eq:we-phichi-2} would leave one with $D{+}1$ equations, whose characteristic polynomial is identically zero; the corresponding matrix has two eigenvalues $\pm [(u{\cdot}\xi)^2 - c^2(\phi) \xi{\cdot}\Delta{\cdot}\xi]^{1/2}$, as well as $D{-}1$ vanishing eigenvalues. This is not surprising: the $D{+}1$ equations \eqref{eq:we-phichi-1}, \eqref{eq:we-phichi-3} contain only two dynamical equations, while the other $D{-}1$ equations are constraints.
In order to isolate the dynamical equations, let us prescribe initial conditions on a spacelike hypersurface~$\Sigma_0$. At each point on $\Sigma_0$, there is a timelike unit co-vector $n_\mu$ normal to $\Sigma_0$, so that $P_{\mu\nu} = g_{\mu\nu} + n_\mu n_\nu$ projects onto $\Sigma_0$. The system of equations \eqref{eq:we-phichi} may then be separated into dynamical equations (those that contain derivatives along $n_\mu$) and constraint equations (those with only derivatives along $\Sigma_0$) at each point on $\Sigma_0$. The system~\eqref{eq:we-phichi} contains $D{+}1$ dynamical equations, and $D(D{-}1)/2$ constraint equations. 

The $D{+}1$ dynamical equations may be written as $M^\mu_{AB}(U)\partial_\mu U^B + N_{A}(U) = 0$, where $U^B = (\chi^\beta, \phi)$. The characteristic form $Q(\xi) = \det(M^\mu \xi_\mu)$ is straightforward to evaluate, and the characteristics are determined by
\begin{equation}
\label{eq:Qw11}
    Q(\xi) = (-\det g) \lr{n{\cdot}\xi}^{D-1} \lr{\lr{u{\cdot}\xi}^2 - c^2(\phi) \Delta^{\mu\nu}\xi_\mu \xi_\nu} = 0\,.
\end{equation}
The characteristic equation  \eqref{eq:Q11} of the original second order wave equation is still present, however equation~\eqref{eq:Qw11} also has $D{-}1$ solutions which correspond to characteristics with normal co-vectors wholly contained within the initial hypersurface $\Sigma_0$. These correspond to non-propagating solutions of the first-order system of dynamical equations which do not solve the original second-order wave equation~\eqref{eq:ex-1}. Importantly, the very separation of the original first-order system \eqref{eq:we-phichi} into ``dynamical'' and ``constraint'' equations depends on the arbitrarily chosen $n_\mu$.
A different initial hypersurface $\Sigma_0'$ will have a different normal $n'_\mu$, and thus the first-order system with initial conditions on $\Sigma'_0$ will be comprised of a different set of dynamical equations, with a different set of extra characteristics, corresponding to $(n'{\cdot}\xi)=0$. Thus, the first-order dynamical equations by themselves (ignoring the constraint) are not covariant in the sense that the characteristics of the system depend on an arbitrarily chosen~$n$. 

In particular, for initial spacelike hyperplanes, changing from $n_\mu$ to $n'_\mu = \Lambda_{\mu}^{\,\,\,\,\nu} n_\nu$, where $\Lambda_{\mu}^{\,\,\,\,\,\nu}$ is a Lorentz boost, is equivalent to leaving $n_\mu$ unchanged and boosting $u^\mu\to u'^\mu = \Lambda_{\nu}^{\,\,\,\,\mu} u^\nu$, $\xi_\mu \to \xi'_\mu = \Lambda^{\nu}_{\,\,\,\,\mu} \xi_\nu$. 
The solutions to $n{\cdot}\xi=0$ will not be solutions to $n{\cdot}\xi'=0$ after boosting; however, the solutions to $\lr{\lr{u{\cdot}\xi}^2 - c^2(\phi) \Delta^{\mu\nu}\xi_\mu \xi_\nu} = 0$ will still be solutions after boosting, as one might expect. The non-propagating characteristics which solve $(n{\cdot}\xi)=0$ differ from the non-propagating characteristics which solve $(n'{\cdot}\xi)=0$ , therefore ignoring the constraints implies a loss of boost covariance.

\subsection{Causality and dispersion relations}
\label{sec:causality-disprel}

Let us now discuss how causality properties of the theory \eqref{eq:quasilin_gen} may be obtained from dispersion relations of plane waves in a linearized theory. In order to formulate the linearized theory, one starts with constant ``background'' solutions of equations~\eqref{eq:quasilin_gen}, which necessarily must satisfy ${\cal C}^A[U]=0$; call these solutions $\tilde U^A_0$. 
Linearizing about the background, $U^A(x) = \tilde{U}^A_0 + \delta U^A(x)$, the perturbations satisfy
\begin{equation}\label{eq:linearized_quasi}
    {\cal A}^{\mu\nu}_{AB}[\tilde{U}_0,0] \de_\mu \de_\nu \delta U^B + {\cal B}^{\mu}_{AB}[\tilde{U}_0,0] \de_\mu \delta U^B + {\cal C}'_{AB}[\tilde{U}_0]\delta U^B = 0 \,,
\end{equation}
where ${\cal C}'_{AB} = \partial {\cal C}^A[U] /\partial U^B$. 
Plane waves $\delta U^B(x) = \delta U^B(K) \exp\lr{i K_\mu x^\mu}$ solve the linear equations \eqref{eq:linearized_quasi} provided 
\begin{equation}\label{eq:spectral_curve}
   F(K) \equiv \det\lr{-{\cal A}^{\mu\nu} [\tilde{U}_0,0] K_\mu K_\nu + i {\cal B}^{\mu} [\tilde{U}_0,0]K_\mu + {\cal C}'[\tilde{U}_0]} = 0
\end{equation}
is satisfied. For $K_\mu = \lr{-\omega, \k}$, solving $F(K)=0$ gives rise to dispersion relations $\omega_\ell(\k)$, where the index $\ell$ labels the solutions.
In a covariantly stable linearized theory, the dispersion relations must obey~\cite{Hoult:2023clg}
\begin{equation}
\label{eq:lin_caus_conds}
    \lim_{|\k| \to \infty} \frac{|\Re(\omega_\ell(\k))|}{|\k|} \leq 1, \quad \lim_{|\k| \to \infty} \frac{\Im(\omega_\ell(\k))}{|\k|} = 0, \quad {\cal O}_{\omega}(F(\omega,\k)) = {\cal O}_{|\k|}(F(\omega {=} a |\k|, \k_j {=} s_j |\k|)) \,.
\end{equation}
If these conditions are satisfied, $V'_\mu \equiv K_\mu/|\k|$ is finite in the limit $|\k| \to \infty$, and thus the leading-order large-$\k$ dispersion relations are determined by 
\begin{equation}
\label{eq:lin_F_large_k}
   \det\lr{{\cal A}^{\mu\nu} [\tilde{U}_0,0]V'_\mu V'_\nu} +...= 0 \,,
\end{equation}
where the $...$ terms are sub-leading in $1/|\k|$. To leading order in large-$\k$, this is simply $Q_{\rm lin}$, the characteristic equation for the linearized system of partial differential equations, expressed in terms of $V'_\mu$ rather than $\xi_\mu$. Thus the causality conditions obtained from the leading large-$\k$ dispersion relations in the linearized theory will match the causality conditions obtained from the characteristic equation of the full non-linear theory if equation~\eqref{eq:lin_F_large_k} coincides with the full characteristic equation \eqref{eq:char_eq} upon taking $V'_\mu \to V_\mu$, $\tilde U_0 \to U(x)$. In order for that to happen, it is sufficient to demand that: 
\begin{itemize}
\item[a)] The coefficients ${\cal A}^{\mu\nu}_{AB}$ in equations~\eqref{eq:quasilin_gen} do not depend on $\partial U$, and
\item[b)] The homogeneous solutions $\tilde{U}_0$ of equations~\eqref{eq:quasilin_gen} are not subject to algebraic constraints, apart from those which apply to $U(x)$.
\end{itemize}
To summarize, if one has a quasilinear system of equations of the form \eqref{eq:quasilin_gen} which satisfy the above conditions a) and b), then demanding that the conditions \eqref{eq:lin_caus_conds} hold for dispersion relations in the linearized system is equivalent to demanding that conditions \eqref{eq:non-linear_caus_conds_xi} hold for the non-linear system. In other words, if the conditions a) and b) are satisfied, one may be assured of the causality of the non-linear system solely by computing the leading-order large-$|\k|$ behaviour of dispersion relations and ensuring they satisfy \eqref{eq:lin_caus_conds}.

While the above conditions a) and b) are sufficient, they are not necessary: as illustrated in section~\ref{sec:examples}, it is possible to violate both of them, and still have the linearized theory correctly predict non-linear causality; this would happen if the violations of a) and b) do not make their way into the characteristic equation. 

The condition a) is a property that a given theory \eqref{eq:quasilin_gen} simply does or does not have; on the other hand, condition b) can be a matter of setting up the linearization. In particular, among other things condition b) says that for non-zero $U^A$, all components of $\tilde U^A_0 = (\tilde U^1_0, \dots, \tilde U^n_0)$ must be non-zero. As was also illustrated in section~\ref{sec:examples}, in some cases it is possible to use the linearized theory in order to correctly assess the non-linear causality properties, even if the violations of b) do enter the characteristic equation of the linearized theory. 

The latter situation is exactly what is realized in dissipative hydrodynamic theories discussed in references~\cite{Kovtun:2019hdm, Bemfica:2019knx}: a linearized theory can be set up by expanding about the equilibrium state of the fluid at rest, where the fluid velocity takes the form $\tilde u^\mu_0 = (1, {\bf 0})$. While this choice of the background clearly violates the above condition b) due to the algebraic condition on the spatial velocity ($\mathbf{v}_0 = 0$), Lorentz covariance of the principal part ensures that the causality conditions obtained in the linearized theory do match the causality conditions of the full non-linear theory; moreover, the properties \eqref{eq:lin_caus_conds} of the large-$\k$ dispersion relations of the theory linearized about the fluid rest frame can be used to establish the causality properties of the full non-linear theory. 

\subsection{Constraint equations}
\label{sec:constraints}
In the previous subsection, there was an underlying assumption that all of the equations contributed to the evolution of the variables $U^B$. It may instead be the case that only some of the equations describe the time evolution of the variables, while the remainder serve as constraints on initial data, as in \nameref{ex:ex7}.

Let us consider a foliation of spacetime by spacelike hypersurfaces $\Sigma_t$ labelled by a parameter $t$. At each point $p$ on $\Sigma_t$, the normal to the hypersurface is given by a unit timelike covector $n_\mu$; the projector onto the hypersurface is  $P_{\mu\nu} = n_\mu n_\nu + g_{\mu\nu}$. Take $\Sigma_0 = \Sigma_{t=0}$ to be the ``initial hypersurface" upon which we prescribe initial conditions. For second-order systems of partial differential equations such as \eqref{eq:quasilin_gen}, initial conditions are comprised of $U^B \vert_{\Sigma_0}$ and $n^\mu \de_\mu U^B \vert_{\Sigma_0}$. ``Dynamical equations" are then those equations that contain $n^\mu n^\nu \de_\mu \de_\nu U^B$. Equations which do not contain the second normal derivative of $U^B$ are ``constraint equations"; instead of describing the evolution of $U^B$, they constrain the form of the initial data.

Let us suppose that the full system of equations is covariant. The dynamical equations are given by 
\begin{equation}\label{eq:dynamical_gen}
    \tilde {\cal A}^{\mu\nu}_{AB}[U,\de U] \de_\mu \de_\nu U^B + \tilde {\cal B}^\mu_{AB}[U,\de U] \de_\mu U^B + \tilde {\cal C}_A[U] = 0\,,
\end{equation}
where $\det(\tilde {\cal A}^{\mu\nu}_{AB}n_\mu n_\nu)$ must be non-vanishing.
The dynamical equations \eqref{eq:dynamical_gen} are supplemented by $N_{\rm c}$ constraint equations of the form
\begin{equation}\label{eq:constraint_gen}
    \tilde {\cal D}^{\mu\nu}_{a B}[U,\de U]\de_\mu \de_\nu U^B + \tilde {\cal E}^\mu_{a B}[U, \de U] \de_\mu U^B + \tilde {\cal F}_{a}[U] = 0\,,
\end{equation}
where $a=1,\dots, N_{\rm c}$ labels constraint equations and $\tilde {\cal D}^{\mu\nu}_{a B} n_\mu n_\nu = 0$. The full system \eqref{eq:dynamical_gen}, \eqref{eq:constraint_gen} is assumed to be covariant, but the decomposition of the system into dynamical and constraint equations is not -- it is instead dependent on how one defines the hypersurfaces $\Sigma_t$, and the normal vector $n_\mu$. In particular, this means that \eqref{eq:dynamical_gen} by itself is not covariant, and characteristics will be sensitive to a choice of $n$. 
The only exception is if $\tilde {\cal D} = 0$. 

Let us consider a system in which both the dynamical and constraint equations satisfy the sufficient conditions a) and b) of the previous subsection, namely
\begin{subequations}
\label{eq:dyn_con_sufficient}
\begin{align}
    \tilde {\cal A}^{\mu\nu}_{AB}[U] \de_\mu \de_\nu U^B + \tilde {\cal B}_{AB}^\mu[U,\de U] \de_\mu U^B &= 0\,,\label{eq:dyn_sufficient}\\
    \tilde {\cal D}^{\mu\nu}_{a B}[U]\de_\mu \de_\nu U^B + \tilde {\cal E}^\mu_{a B}[U, \de U] \de_\mu U^B &= 0\,.
\end{align}
\end{subequations}
The equivalence between linear and non-linear causality%
\footnote{
\label{fn:cc2}%
Here, by ``causality" we mean that for any point $p \in \Sigma_t$, the characteristics of the dynamical equations at $p$ are entirely within the lightcone, for any $n_\mu$.
} 
holds for such a system, even in the presence of constraints.
However, because of the constraint equations, Lorentz transformations are not a symmetry of the principal part of equation~\eqref{eq:dyn_sufficient} anymore. In the notation of equation~\eqref{eq:A-symmetry}, where $G$ corresponds to Lorentz transformations, Lorentz invariance of the characteristic equation would require $G \tilde {\cal A}(\phi,n) G^{-1} = \tilde {\cal A}(G \phi, n)$, which is not true. 
In the context of relativistic hydrodynamics, this means that it is not sufficient to ensure linearized causality purely for the fluid at rest, with $\tilde u^\mu_0 = (1, {\bf 0})$; one must ensure causality in all reference frames.
As such, one must look at linearized fluctuations about a general $\tilde{U}_0^A$. 

This is an annoying property for the system to have, and so it would be preferable to find a way to utilize the constraint equations to restore covariance to the dynamical equations. 
We plan to return to the question of restoring covariance in future work. For now, we simply note that in those cases (such as in \nameref{ex:ex7}) where the characteristic equation for the dynamical equations is of the form%
\footnote{
   When the second-order equations of BDNK hydrodynamics \cite{Bemfica:2017wps, Kovtun:2019hdm, Bemfica:2019knx, Hoult:2020eho} are reduced to first order, the characteristic equation of the resulting first-order system with constraints has the form \eqref{eq:good_form_characteristic}. Dissipative relativistic magnetohydrodynamics discussed later in section~\ref{sec:intro-MHD} is a system of second-order equations with constraints; the characteristic equation again has the form~\eqref{eq:good_form_characteristic}, see section~\ref{sec:nlc}. 
}
\begin{equation}
\label{eq:good_form_characteristic}
    Q = \lr{n{\cdot}\xi}^\ell \tilde{Q} = 0 \,,
\end{equation}
where $\tilde{Q}$ is a Lorentz scalar independent of $n$, and $\ell>0$ is a positive integer, the non-covariance is entirely contained in $n{\cdot}\xi$. Since the solutions to $(n{\cdot}\xi) = 0$ are always causal for any timelike $n_\mu$, causality of the system comes down to determining the roots of $\tilde{Q}$. Since $\tilde{Q}$ is a Lorentz scalar, the equation $\tilde{Q} = 0$ may be solved with any choice of $n_\mu$.

When analyzing $\tilde Q$ in a hydrodynamic theory, one can choose to work in the local fluid rest frame. As $n_\mu$ defines the coordinate time, the fluid rest frame corresponds to $n{\cdot}u=-1$ at a given point $p$. After fixing this choice of $n_\mu$, the only non-zero component of $u$ at $p$ (in the non-linear theory) is the component along $n$, i.e.\ $\mathbf{v} = 0$. Then, performing the linearized analysis about a constant solution $u^\mu_0 = (1, \mathbf{0})$ does not violate the sufficient condition b), as the algebraic constraint $\mathbf{v}_0 = 0$ is simply inherited from $u^\mu(x)$. The sufficient conditions for the linear vs non-linear equivalence are then satisfied (for that particular choice of $n_\mu$, and $p$). One can thus find the roots of $\tilde{Q}=0$ from linearized fluctuations about the fluid rest frame, and determine the conditions for non-linear causality. Then, the Lorentz-invariance of $\tilde{Q}$ implies that the causality conditions (such as constraints on the fluid transport parameters in first-order hydrodynamics) obtained in the fluid rest-frame are in fact sufficient to render the non-linear system causal for any $n_\mu$.

\subsection{Hydrodynamics plus gravity}
Let us now discuss first-order dissipative hydrodynamic theories with no derivative constraints coupled to gravity. 
We denote the hydrodynamic variables such as temperature, fluid velocity etc.\ collectively as $\psi^a$, with $a=1, \dots, M$. The equations are
\begin{align}
\label{eq:fluid-GR}
  \nabla_{\!\mu} T^{\mu\nu} = 0 \,,\ \ \ \ 
  \nabla_{\!\mu} J^{\mu}_i = 0\,,\ \ \ \ 
  G_{\mu\nu} = 8\pi G_{\rm N} T_{\mu\nu} \,,
\end{align}
where $\nabla_{\!\mu}$ is the covariant derivative, $T^{\mu\nu}$ is the energy-momentum tensor, $J^\mu_i$ are conserved currents for ``species'' labeled by~$i$, $G_{\mu\nu}$ is the Einstein tensor, and $G_{\rm N}$ is Newton's constant. In first-order hydrodynamics, the conserved currents are given in terms of hydrodynamic variables $\psi$ and the metric $g$; schematically, $T^{\mu\nu} = T^{\mu\nu}_{\rm p.f.}(\psi, g) + a(\psi, g) O(\partial \psi) + b(\psi, g) O(\partial g)$, and similarly for $J^\mu_i$. The first term $T^{\mu\nu}_{\rm p.f.}(\psi, g)$ describes ``perfect fluids'', the terms proportional to $O(\partial \psi)$ encode transport parameters such as viscosities and conductivities, and the terms with $O(\partial g)$ arise in curved space due to contributions such as $\nabla_{\!\mu} u_\nu$ in the energy-momentum tensor. Importantly, there are no product terms of the form $(\partial \psi) (\partial g)$ in either $T^{\mu\nu}$ or $J^\mu_i$. Introducing the vector of variables $U^A = (\psi^a, g_I)$, the theory \eqref{eq:fluid-GR} can be cast in the form \eqref{eq:quasilin_gen}. Here the index $I$ labels the metric components,  in $D$ spacetime dimensions $I =1, \dots, N$, with $N = D(D{+}1)/2$. The principal part of \eqref{eq:quasilin_gen} then takes the form \cite{Bemfica:2019knx, Hoult:2020eho}
\begin{equation}
\label{eq:fluid-GR-pp}
  {\cal A}^{\mu\nu}_{AB}[U] \partial_\mu \partial_\nu U^B = \begin{bmatrix}
{\cal X}^{\mu\nu}[\psi,g]_{ab} & {\cal Y}^{\mu\nu}[\psi,g]_{aJ}\\
0_{N\times M} & g^{\mu\nu} \, \delta_{IJ}
\end{bmatrix}
\partial_\mu \partial_\nu 
  \begin{pmatrix}
    \psi^b\\
    g_{J}
  \end{pmatrix} \,.
\end{equation}
The upper-left ${\cal X}^{\mu\nu}[\psi,g]$ is a $M\times M$ matrix, the upper-right ${\cal Y}^{\mu\nu}[\psi, g]$ is a $M\times N$ matrix. The bottom-right block reflects the Einstein equations written in the harmonic gauge as $g^{\mu\nu} \partial_\mu \partial_\nu g_{\alpha\beta} + \dots = 0$. The bottom-left block is zero. Thus, in first-order hydrodynamics, the coefficient ${\cal A}^{\mu\nu}[U]$ of the principal part does not depend on derivatives of the functions. The block structure of the principal part \eqref{eq:fluid-GR-pp} implies that the characteristic equation factorizes into a product of the ``fluid part'' and the ``gravity part''. The latter, $g^{\mu\nu} \xi_\mu \xi_\nu = 0$, defines the light-cone, while the former, 
\begin{align}
\label{eq:fluid-GR-cf}
  Q_{\rm fluid} \equiv \det \left( {\cal X}^{\mu\nu}[\psi, g]\, \xi_\mu \xi_\nu \right) = 0\,,
\end{align}
needs to be solved for $\xi_\mu$. Imposing the causality conditions \eqref{eq:non-linear_caus_conds_xi} on the solutions of \eqref{eq:fluid-GR-cf} will impose causality constraints on the transport parameters of the hydrodynamic theory. The characteristic equation \eqref{eq:fluid-GR-cf} is local and diffeomorphism-invariant. Therefore, it can be expressed in terms of local diffeomorphism invariants which are quadratic in $\xi$. For example, for a hydrodynamic theory whose degrees of freedom comprise temperature $T$, fluid velocity $u^\mu$, and scalar chemical potentials $\mu_i$, the characteristic equation will be of the form $F(\zeta_1, \zeta_2 )  = 0$, where $\zeta_1 \equiv u^\mu \xi_\mu$, $\zeta_2  \equiv (g^{\mu\nu} {+} u^\mu u^\mu)\xi_\mu \xi_\nu$, and  $F(\zeta_1, \zeta_2)$ is a polynomial whose coefficients depend on $T$ and $\mu_i$ (through the equation of state and the transport parameters). Locally, at a given point $x_p$, one can pass to Gaussian normal coordinates, so that the metric is flat at that point, and $g^{\mu\nu}(x_p) = \eta^{\mu\nu}$. Thus, the characteristic equation at $x_p$ will be that of a fluid in flat space.
The causality conditions obtained from the characteristic equation are thus the causality conditions of a fluid in flat space; the causality constraints are expressed in terms of the equation of state and the transport parameters, which are local functions of $T(x_p)$ and $\mu_i(x_p)$. As described in section~\ref{sec:causality-disprel}, in order to find the causality conditions for fluid in flat space, it suffices to perform a linearized analysis of the fluid equations, expanded about the state with constant $T$, $\mu_i$, and $u^\mu = (1, {\bf 0})$. The large-$\k$ dispersion relations in this linearized theory will then give the correct causality conditions for non-linear dissipative hydrodynamics coupled to dynamical gravity.

\section{Magnetohydrodynamics}
\label{sec:intro-MHD}
In the previous section, we have discussed how causality conditions of the non-linear theory \eqref{eq:quasilin_gen} can be obtained from the analysis of large-$\k$ dispersion relations in the linearized theory. The main ingredient which allowed the linearized causality analysis to work was the requirement that the coefficient ${\cal A}^{\mu\nu}$ of the system \eqref{eq:quasilin_gen} not depend on the derivatives. In order to demonstrate the usefulness of this equivalence between linear and non-linear causality, we will now look at a system which has not previously been subject to a non-linear causality analysis, dissipative relativistic magnetohydrodynamics formulated using the conservation of magnetic flux (dMHD). We include a conserved neutral-particle current $J^\mu$, in order to facilitate possible astrophysical applications.%
\footnote{
  See e.g.~\cite{Gammie:2003rj, DelZanna:2007pk, Beckwith:2011iy, Porth:2016rfi, Mignone:2019ebw} for numerical implementations of relativistic MHD supplemented by conservation of $J^\mu$ for astrophysical applications. 
}
In this section, we will write down the dMHD equations; in the next section, we will study their causality properties.

\subsection{Perfect-fluid MHD}
We start with perfect-fluid relativistic MHD in a fixed background spacetime. The hydrodynamic variables are the temperature $T$, fluid velocity $u^\mu$, and the magnetic field $B^\mu$, defined in $3+1$ dimensions in terms of the electromagnetic field strength $F_{\mu\nu}$ as $B^\mu = \frac12 \epsilon^{\mu\nu\rho\sigma} u_\nu F_{\rho\sigma}$, so that $u{\cdot}B = 0$. The Levi-Civita tensor is $\epsilon^{\mu\nu\rho\sigma} = \varepsilon^{\mu\nu\rho\sigma}/\sqrt{-g}$, where $\varepsilon$ is the antisymmetric symbol with components $0, \pm1$. For fluids with an extra conserved global $U(1)$ charge, such as conserved particle number of electrically neutral particles, the corresponding chemical potential $\mmu$ is also a hydrodynamic variable.  The hydrodynamic equations are the conservation of the energy-momentum tensor $T^{\mu\nu}$ and of the current $J^\mu$, plus that half of Maxwell's equations which determines the dynamics of the magnetic field,%
\begin{subequations}
\label{eq:TJJJ-conserv}
\begin{align}
\label{eq:T-conserv}
  & \nabla_{\!\mu} T^{\mu\nu} = 0 \,,\\
\label{eq:JJ-conserv}
  & \nabla_{\!\mu} J^{\mu\nu} = 0 \,,\\
\label{eq:J-conserv}
  & \nabla_{\!\mu} J^\mu = 0\,,
\end{align}
\end{subequations}
where $J^{\mu\nu} = \frac12 \epsilon^{\mu\nu\rho\sigma} F_{\rho\sigma}$ is the dual electromagnetic field strength. Electric fields (in the fluid rest frame) $E_\mu = F_{\mu\nu} u^\nu$ are assumed to vanish due to electric screening. In 3+1 dimensions, equations~\eqref{eq:TJJJ-conserv} are $8$ equations for $8$ variables $T$, $\mmu$, $u^\mu$ (constrained so that $u^2 = -1$), and $B^\mu$ (constrained so that $u{\cdot}B = 0$). The time component of equation~\eqref{eq:JJ-conserv} is not a dynamical equation, but rather a constraint on the magnetic field (no magnetic monopoles). 

In order to write down the hydrodynamic equations, the conserved quantities $T^{\mu\nu}$, $J^{\mu\nu}$, and $J^\mu$ in \eqref{eq:TJJJ-conserv} need to be expressed in terms of the hydrodynamic variables $T$, $\mmu$, $u^\mu$, and $B^\mu$. The guidance for how to do so comes from equilibrium. In perfect-fluid hydrodynamics, the expressions for conserved currents are first written down in global equilibrium, e.g.\ $J^\mu = n u^\mu$, where $n$ is the equilibrium charge density, and $u^\mu$ is the constant fluid velocity. Next, the same equilibrium expressions are postulated to describe the non-equilibrium evolution, when used inside time-dependent conservation laws, e.g.\ $\nabla_{\!\mu} (n u^\mu) = 0$ is assumed to be the correct dynamical equation for time-dependent $n$ and $u^\mu$ out of equilibrium. This standard hydrodynamic assumption can be used to write down the equations of perfect-fluid MHD. 

The expressions for $T^{\mu\nu}$ and $J^\mu$ can be obtained from the equilibrium hydrostatic effective action $S_{\rm eff}[g, \A, A] = \int\!\sqrt{-g} \, {\cal F}$, where $g$ is the static external metric, $\A$ is the static external gauge field coupled to $J^\mu$, and $A$ is the electromagnetic gauge field in equilibrium. The action density is ${\cal F}(T, \mmu, B^2)$, where $T$ and $\mmu$ depend on $g$ and $\A$ through the standard hydrostatic expressions, in particular, $T = T_0/(- g_{\mu\nu} V^\mu V^\nu)^{1/2}$ is the Tolman temperature, where $V$ is the timelike Killing vector which specifies the equilibrium state, and the constant $T_0$ sets the overall normalization of temperature. See e.g.\ references~\cite{Hernandez:2017mch, Kovtun:2016lfw} for more details. In the perfect-fluid approximation, electric fields are neglected, hence ${\cal F}$ does not depend on either $E^2$ or $E{\cdot}B$. Further, local  charge neutrality for electric charges ensures that ${\cal F}$ does not depend on the chemical potential for the conserved electric charge. It is customary to isolate the Maxwell part of the action, and write ${\cal F}(T, \mmu, B^2) = p_{\rm m}(T, \mmu, B^2) - \frac12 B^2$, where $p_{\rm m}$ is the ``matter'' pressure. The magnetic permeability is $\mu_{\rm B}(T, \mmu, B^2) = 1/(1- 2\partial p_{\rm m}/\partial B^2)$. 

Varying the above $S_{\rm eff}[g, \A, A]$ with respect to $g_{\mu\nu}$ gives rise to $T^{\mu\nu}$, varying with respect to $\A_\mu$ gives rise to $J^\mu$, and varying with respect to $A_\mu$ gives rise to Maxwell's equations in matter. The resulting $T^{\mu\nu}$ and $J^\mu$ are automatically conserved in hydrostatic equilibrium, thanks to diffeo- and gauge-invariance of $S_{\rm eff}$. The conserved quantities in equation~\eqref{eq:TJJJ-conserv} are then expressed in terms of a single function $p_{\rm m}(T, \mmu, B^2)$, the grand canonical pressure of matter, as%
\begin{subequations}
\label{eq:pfMHD}
\begin{align}
\label{eq:pfMHD-T}
  & T^{\mu\nu} = \left(  T \frac{\partial p_{\rm m}}{\partial T} + \mmu \frac{\partial p_{\rm m}}{\partial \mmu} + \frac{B^2}{\mu_{\rm B}} \right) u^\mu u^\nu + \left( p_{\rm m} - \coeff12 B^2 + \frac{B^2}{\mu_{\rm B}} \right) g^{\mu\nu} - \frac{B^\mu B^\nu}{\mu_{\rm B}} \,, \\
\label{eq:pfMHD-JJ}
  & J^{\mu\nu} = u^\mu B^\nu - u^\nu B^\mu \,,\\
\label{eq:pfMHD-J}
  & J^{\mu} = \left( \frac{\partial p_{\rm m}}{\partial \mmu} \right) u^\mu \,.
\end{align}
\end{subequations}
The functional form of $p_{\rm m}(T, \mmu, B^2)$ is provided by the equation of state in the grand canonical ensemble. If the effects of magnetic polarization in matter are neglected, then $p_{\rm m} = p_{\rm m}(T, \mmu)$, and $\mu_{\rm B} = 1$.  The density of neutral particles is $n_{\rm m} \equiv \partial p_{\rm m}/\partial\mmu$, and the entropy density is $s_{\rm m} \equiv \partial p_{\rm m}/\partial T$. The corresponding enthalpy density of matter is $w_{\rm m} \equiv T s_{\rm m} + \mmu\, n_{\rm m}$, and the energy density of matter is $\epsilon_{\rm m} \equiv -p_{\rm m} + T s_{\rm m} + \mmu\,  n_{\rm m}$. 

As an example, when the constituents of $J^\mu$ are massive neutral particles of mass $m$, one can work with a ``mass current'' $J^\mu_m \equiv m J^\mu$ instead of $J^\mu$. The mass density is then $\rho_{\rm m} \equiv m n_{\rm m}$, and the internal energy density is $(\epsilon_{\rm m} - \rho_{\rm m})$. For the ideal gas equation of state $p_{\rm m} = (\gamma {-} 1) (\epsilon_{\rm m} - \rho_{\rm m})$, where $\gamma$ is the adiabatic index, the conserved currents are
\begin{subequations}
\label{eq:pfMHD-3}
\begin{align}
  & T^{\mu\nu} = \left(  \rho_{\rm m} + \coeff{\gamma}{\gamma-1} p_{\rm m} + B^2 \right) u^\mu u^\nu + \left( p_{\rm m} + \coeff12 B^2  \right) g^{\mu\nu} - B^\mu B^\nu \,, \\
  & J^{\mu\nu} = u^\mu B^\nu - u^\nu B^\mu \,,\\
  & J^{\mu}_m = \rho_{\rm m} u^\mu \,.
\end{align}
\end{subequations}
Regardless of the equation of state used, \eqref{eq:TJJJ-conserv} - \eqref{eq:pfMHD} are the equations of perfect-fluid MHD. They will be causal so long as the speeds of magnetosonic and Alfv\'en waves are subluminal.

\subsection{Perfect-fluid dMHD}
\label{sec:dMHD_setup}

In the above presentation of MHD, the conservation equations for energy, momentum, and particles stood on a conceptually different footing compared to the equations which govern the dynamics of the magnetic field (Faraday's law plus the no-monopole condition). While \eqref{eq:T-conserv} reflects the spacetime symmetry and \eqref{eq:J-conserv} reflects a global $U(1)$ symmetry, equation~\eqref{eq:JJ-conserv} is not usually presented as a manifestation of symmetry in conventional discussions of MHD. As a result, while the constitutive relations \eqref{eq:pfMHD-T} and \eqref{eq:pfMHD-J} are written on the basis of equilibrium thermodynamics, equation~\eqref{eq:pfMHD-JJ}, as written, is not based on any thermodynamics: after all, in perfect-fluid MHD, equation~\eqref{eq:pfMHD-JJ} is just a consequence of the definition of $B^\mu$. 

There {\em is}, however, a global symmetry which is responsible for equation~\eqref{eq:JJ-conserv}, a ``one-form'' symmetry of electromagnetism which reflects the conservation of magnetic flux. It is an example of a generalized symmetry in gauge theories, see reference~\cite{Gaiotto:2014kfa} for a discussion of such symmetries. The usefulness of this one-form symmetry for MHD was emphasized in reference~\cite{Grozdanov:2016tdf}, see also \cite{Schubring:2014iwa} for earlier related work. In practice, using the one-form symmetry for perfect-fluid MHD amounts to viewing equation~\eqref{eq:JJ-conserv} as conservation of the current $J^{\mu\nu}$, for which one needs to provide constitutive relations in terms of hydrodynamic variables, order by order in the derivative expansion, exactly as is done for $T^{\mu\nu}$ and $J^\mu$.  The conserved magnetic flux density can be viewed as a thermodynamic variable $\rhophi$ coming with an associated chemical potential $\muphi$, see references~\cite{Grozdanov:2016tdf, Armas:2018atq} for details of the thermodynamic construction. The direction of the magnetic field lines is parametrized by a unit spacelike vector $h^\mu$. When thermodynamic variables are promoted to hydrodynamic variables out of equilibrium,  the constitutive relations in the magnetohydrodynamic theory (which we term dMHD) are to be written in terms of $T, u^\alpha, \mmu, \muphi, h^\alpha$ (and their derivatives if one would like to go beyond perfect fluids). 

The perfect-fluid constitutive relations in the one-form formulation of MHD are expressed in terms of the pressure function $p = p(T, \mmu, \muphi)$ as%
\begin{subequations}
\label{eq:pfMHD-dual}
\begin{align}
\label{eq:pfMHD-dual-T}
  & T^{\mu\nu} = \left(  T \frac{\partial p}{\partial T} + \mmu \frac{\partial p}{\partial \mmu} + \muphi \frac{\partial p}{\partial \muphi}  \right) u^\mu u^\nu + p g^{\mu\nu} - \muphi \rhophi h^\mu h^\nu \,,\\
\label{eq:pfMHD-dual-JJ}
  & J^{\mu\nu} = \rhophi (u^\mu h^\nu - u^\nu h^\mu) \,, \\
\label{eq:pfMHD-dual-J}
  & J^{\mu} = \left( \frac{\partial p}{\partial \mmu} \right) u^\mu \,,
\end{align}
\end{subequations}
where $\rhophi \equiv \partial p/\partial\muphi$. The constitutive relations \eqref{eq:pfMHD-dual} of dMHD are in fact the same as the corresponding constitutive relations \eqref{eq:pfMHD} of the ``conventional'' MHD~\cite{Hernandez:2017mch}. 
Defining $B\equiv \sqrt{B^2}$, one can relate the thermodynamic parameters between \eqref{eq:pfMHD} and \eqref{eq:pfMHD-dual} as follows: $h^\mu = B^\mu /B$, $\rhophi = B$, $\muphi = B/\mu_{\rm B}$ (alternatively, $\muphi = -2B \, \partial {\cal F}/\partial B^2$), and $p = p_{\rm m} - B^2/2 + {B^2}/{\mu_{\rm B}}$. Note that $(\partial p/\partial T)_{\mmu, \muphi} = (\partial p_{\rm m}/\partial T)_{\mmu, B}$, and also $(\partial p/\partial \mmu)_{T, \muphi} = (\partial p_{\rm m}/\partial \mmu)_{T, B}$. Defining the energy density $\epsilon \equiv -p + T(\partial p/\partial T) + \mmu (\partial p/\partial \mmu) + \muphi (\partial p/\partial \muphi)$, one then finds $\epsilon = -{\cal F} + T (\partial {\cal F}/\partial T) + \mmu (\partial {\cal F}/\partial \mmu)$, or $\epsilon = \epsilon_{\rm m} + \coeff12 B^2$. The number density $n\equiv \partial p/\partial \mmu$, coincides with $n_{\rm m} = \partial p_{\rm m}/\partial\mmu$. Similarly, the entropy density $s\equiv \partial p/\partial T$ coincides with $s_{\rm m} = \partial p_{\rm m}/\partial T$. The thermodynamic relation $dp = sdT + n d\mmu + \rhophi d \muphi$ can be written as $dp = sdT + n d\mmu + B d H$, where $B$ is the magnetic flux density, and $H$ is the magnetic $H$-field. The pressure function is $p = {\cal F} + HB$.

The conservation equations \eqref{eq:TJJJ-conserv} are to be solved together with the constitutive relations \eqref{eq:pfMHD-dual} to find the hydrodynamic fields $T$, $u^\mu$, $\mmu$, $\muphi$, and $h^\mu$. As an example, let us take matter with no magnetic polarization, $\mu_{\rm B} =1$, so that $\rhophi(T,\mmu, \muphi) = \muphi$, and work with the ``mass current'' $J^\mu_m = m J^\mu$. The perfect-fluid constitutive relations \eqref{eq:pfMHD-dual} with the ideal-gas equation of state $p_{\rm m} = (\gamma {-} 1) (\epsilon_{\rm m} - \rho_{\rm m})$ take the form
\begin{subequations}
\label{eq:pfMHD-dual-3}
\begin{align}
  & T^{\mu\nu} = \left(  \rho_{\rm m} + \coeff{\gamma}{\gamma-1} \left( p - \coeff12 \rhophi^2 \right) + \rhophi^2 \right) u^\mu u^\nu + p \,  g^{\mu\nu} - \rhophi^2 \, h^\mu h^\nu \,, \\
  & J^{\mu\nu} = \rhophi ( u^\mu h^\nu - u^\nu h^\mu ) \,,\\
  & J^{\mu}_m = \rho_{\rm m} u^\mu \,.
\end{align}
\end{subequations}
The conservation equations for \eqref{eq:pfMHD-dual} and \eqref{eq:pfMHD-dual-3} are causal so long as the speeds of sound for Alfv\'en and magnetosonic waves in the system are less than the speed of light. 

\subsection{Tensor decompositions}

Before we write down the dissipative corrections to perfect-fluid dMHD, let us look at the general form of conserved currents. Given a timelike unit vector $u^\mu$, a symmetric tensor $T^{\mu\nu}$, an anti-symmetric tensor $J^{\mu\nu}$, and a vector $J^\mu$ can be written as%
\begin{subequations}
\label{eq:TJJJ-general-u}
\begin{align}
\label{eq:TJJJ-general-u-a}
  & T^{\mu\nu} = {\cal E} u^\mu u^\nu + {\cal P} \Delta^{\mu\nu} + {\cal Q}^\mu u^\nu + {\cal Q}^\nu u^\mu + {\cal T}^{\mu\nu}  \,, \\
\label{eq:TJJJ-general-u-b}
  & J^{\mu\nu} =  u^\mu {\cal B}^\nu - u^\nu {\cal B}^\mu + {\cal D}^{\mu\nu}\,,\\
\label{eq:TJJJ-general-u-c}
  & J^\mu = {\cal N} u^\mu + {\cal J}^\mu \,,
\end{align}
\end{subequations}
where $\Delta^{\mu\nu} \equiv g^{\mu\nu} + u^\mu u^\nu $ projects onto the space orthogonal to $u$, and the coefficients in \eqref{eq:TJJJ-general-u} satisfy $u{\cdot}{\cal Q} = u{\cdot}{\cal T} = u{\cdot}{\cal B} = u{\cdot}{\cal D} = u{\cdot} {\cal J} = 0$, ${\cal T}^{\mu\nu} = {\cal T}^{\nu\mu}$, ${\cal D}^{\mu\nu} = -{\cal D}^{\nu\mu}$, $g_{\mu\nu} {\cal T}^{\mu\nu} = 0$. 
The decompositions \eqref{eq:TJJJ-general-u} define%
\footnote{
  Parentheses on indices denote symmetrization, $X_{\dots (\mu \dots} Y_{\dots \nu) \dots} =  (X_{\dots \mu \dots} Y_{\dots \nu \dots} + X_{\dots \nu \dots} Y_{\dots \mu \dots})$, and square brackets denote anti-symmetrization, $X_{\dots [\mu \dots} Y_{\dots \nu] \dots} = (X_{\dots \mu \dots} Y_{\dots \nu \dots} - X_{\dots \nu \dots} Y_{\dots \mu \dots})$.
}
\begin{align}
  & {\cal E} \equiv T^{\mu\nu} u_\mu u_\nu \,,\ \ \ \ 
    {\cal P} \equiv \coeff13 \Delta_{\mu\nu} T^{\mu\nu} \,,\ \ \ \ 
    {\cal N} \equiv - u_\mu J^\mu \,,\\
  & {\cal Q}_\mu \equiv - \Delta_{\mu\rho} u_\sigma T^{\rho\sigma} \,,\ \ \ \ 
    {\cal B}^\mu \equiv J^{\mu\nu} u_\nu \,,\ \ \ \ 
    {\cal J}_\mu \equiv \Delta_{\mu\nu} J^\nu \,, \\
  & {\cal T}_{\mu\nu} \equiv \coeff12 \left( \Delta_{( \mu\alpha} \Delta_{\nu) \beta} - \coeff23 \Delta_{\mu\nu} \Delta_{\alpha\beta} \right) T^{\alpha\beta} \,,\ \ \ \ 
    {\cal D}_{\mu\nu} \equiv \coeff12 \Delta_{[\mu\alpha} \Delta_{\nu]\beta} J^{\alpha\beta} \,.
\end{align}
If, in addition to $u^\mu$, there is also a spacelike unit vector $h^\mu$ which satisfies $u_\mu h^\mu = 0$, the coefficients in \eqref{eq:TJJJ-general-u} may be further decomposed using $\Delta^{\mu\nu}_\perp \equiv g^{\mu\nu} + u^\mu u^\nu  - h^\mu h^\nu$ which projects onto the space orthogonal to both $u$ and $h$. Namely,
\begin{subequations}
\label{eq:TJJJ-general-uh}
\begin{align}
  & T^{\mu\nu} = {\cal E} u^\mu u^\nu + \left( {\cal P} {-} \coeff12 {\cal S} \right) \Delta^{\mu\nu}_\perp + \left( {\cal P} {+} {\cal S} \right) h^\mu h^\nu  + {\cal Q}_\parallel h^{(\mu} u^{\nu)} +  {\cal Q}^{(\mu}_\perp u^{\nu)} + {\cal T}_\perp^{(\mu} h^{\nu)}+ {\cal T}^{\mu\nu}_{\perp} \,, \\
  & J^{\mu\nu} = {\cal B}_\parallel u^{[\mu} h^{\nu]} + u^{[\mu} {\cal B}^{\nu ]}_\perp + {\cal D}^{[\mu}_\perp h^{\nu]} + {\cal D}^{\mu\nu}_\perp \,, \\
  & J^\mu = {\cal N} u^\mu + {\cal J}_\parallel h^\mu + {\cal J}^\mu_\perp \,,
\end{align}
\end{subequations}
where the ``$\perp$'' symbol denotes a quantity that is orthogonal to both $u$ and $h$. The vectors are written as ${\cal Q}^\mu = {\cal Q}_\parallel h^\mu + {\cal Q}^\mu_\perp$, and similarly for ${\cal B}^\mu$ and ${\cal J}^\mu$. The antisymmetric ${\cal D}^{\mu\nu}$ is written as ${\cal D}^{\mu\nu} = {\cal D}^\mu_\perp h^\nu - {\cal D}^\nu_\perp h^\mu + {\cal D}^{\mu\nu}_\perp$, and similarly $  {\cal T}^{\mu\nu} =  \left( h^\mu h^\nu - \coeff12 \Delta^{\mu\nu}_\perp \right) {\cal S}  + {\cal T}^\mu_\perp h^\nu + {\cal T}^\nu_\perp h^\mu + {\cal T}^{\mu\nu}_\perp $. Thus the coefficients which appear in the decomposition \eqref{eq:TJJJ-general-uh} are 
\begin{align}
   & {\cal Q}_\parallel \equiv {\cal Q}^\mu h_\mu \,,\ \ \ \ 
     {\cal Q}^\mu_\perp \equiv \Delta^{\mu\nu}_\perp {\cal Q}_\nu \,, \\
   & {\cal D}^\mu_\perp  \equiv {\cal D}^{\mu\nu} h_\nu \,,\ \ \ \ 
     {\cal D}^{\mu\nu}_\perp \equiv \coeff12 \Delta^{[\mu\alpha}_\perp \Delta^{\nu]\beta}_\perp {\cal D}_{\alpha\beta} \,, \ \ \ \ 
     {\cal S} \equiv h^\alpha h^\beta {\cal T}_{\alpha\beta} \,, \\
   & {\cal T}^\mu_\perp \equiv \Delta^{\mu\alpha}_\perp h^\beta {\cal T}_{\alpha\beta} \,, \ \ \ \ 
     {\cal T}^{\mu\nu}_\perp \equiv \coeff12 \left( \Delta^{\mu\alpha}_\perp \Delta^{\nu\beta}_\perp + \Delta^{\nu\alpha}_\perp \Delta^{\mu\beta}_\perp - \Delta^{\mu\nu}_\perp \Delta^{\alpha\beta}_\perp \right) {\cal T}_{\alpha\beta} \,.
\end{align}
When the decompositions \eqref{eq:TJJJ-general-u} or \eqref{eq:TJJJ-general-uh} are applied to perfect-fluid MHD, the coefficients ${\cal Q}^\mu$, ${\cal D}^{\mu\nu}$, ${\cal J}^\mu$ vanish, and the expressions for ${\cal E}$, ${\cal P}$, ${\cal T}^{\mu\nu}$, ${\cal B}^\mu$, ${\cal N}$ in terms of the hydrodynamic variables can be read off from the constitutive relations. For example, equations~\eqref{eq:TJJJ-general-u} and \eqref{eq:pfMHD} give 
\begin{align}
  & {\cal E} = \epsilon_{\rm m} + \coeff12 B^2 \,,\ \ \ \ 
    {\cal P} = p_{\rm m} - \coeff12 B^2 +\coeff23 \frac{B^2}{\mu_{\rm B}} \,,\ \ \ \ 
    {\cal T}^{\mu\nu} = \frac{1}{\mu_{\rm B}} \left( \coeff13 \Delta^{\mu\nu} B^2 - B^\mu B^\nu \right) \,, \\
  & {\cal B}^\mu = B^\mu\,,\ \ \ \ {\cal N} = n_{\rm m} \,.
\end{align}
Alternatively, combining equations~\eqref{eq:TJJJ-general-uh} and \eqref{eq:pfMHD-dual}, one finds
\begin{align}
  & {\cal E} = \epsilon \,, \ \ \ \ 
    {\cal P} = p - \coeff13 \muphi \rhophi \,, \ \ \ \ 
    {\cal S} = -\coeff23 \muphi \rhophi \,,\\
  & {\cal B}_\parallel = \rhophi \,,\ \ \ \ 
    {\cal N}  = n \,, 
\end{align}
with the rest of the coefficients in \eqref{eq:TJJJ-general-uh} vanishing.

\subsection{Hydrodynamic variables}

The next step is to write down the constitutive relations for dissipative dMHD. In order to do so, one has to choose the set of hydrodynamic variables $U$. The constitutive relations in one-derivative hydrodynamics take the form $T^{\mu\nu} = O(U) + O(\partial U)$, and similarly for other conserved currents. Here we choose to work with the same variables as appear in the perfect-fluid description%
\footnote{
  The choice of hydrodynamic variables is the most important ingredient that goes into writing down a dissipative hydrodynamic theory. As mentioned in the Introduction, the widely used M\"uller-Israel-Stewart-type theories use extra variables, besides those used in perfect-fluid hydrodynamics. 
}
in section~\ref{sec:dMHD_setup}, namely we choose our hydrodynamic variables to be $U=\{ T, u^\alpha, \mmu, \muphi, h^\alpha \}$, where $u^2 = -1$, $h^2 = 1$, and $u{\cdot}h=0$. The scalar variables $T$, $\mmu$, $\muphi$ may be traded for other thermodynamic variables such as $p$, $n$, $\rhophi$, if allowed by the given equation of state. The constitutive relations in dMHD are then%
\begin{subequations}
\label{eq:dMHD-1}
\begin{align}
  & T^{\mu\nu} = T^{\mu\nu}_{(0)} + T^{\mu\nu}_{(1)}  \,, \\
  & J^{\mu\nu} = J^{\mu\nu}_{(0)} + J^{\mu\nu}_{(1)}  \,, \\
  & J^\mu = J^\mu_{(0)} + J^\mu_{(1)} \,,
\end{align}
\end{subequations}
where $T^{\mu\nu}_{(0)}$, $J^{\mu\nu}_{(0)}$, and $J^{\mu}_{(0)}$ are given by the perfect-fluid expressions \eqref{eq:pfMHD-dual} (or by \eqref{eq:pfMHD-dual-3} for the ideal-gas equation of state), and $T^{\mu\nu}_{(1)}$, $J^{\mu\nu}_{(1)}$, and $J^{\mu}_{(1)}$ are one-derivative corrections, linear in derivatives $\partial U$. These one-derivative corrections will contain viscosities and particle-number conductivities (or heat conductivities), as is standard in hydrodynamics~\cite{LL6}. In dMHD, electric resistivities will also appear as transport coefficients. The conservation laws \eqref{eq:TJJJ-conserv} combined with the one-derivative constitutive relations \eqref{eq:dMHD-1} then give the equations of dMHD as quasilinear second-order partial differential equations of the form \eqref{eq:quasilin_gen}. 

Out of equilibrium, the standard hydrodynamic variables $T$, $u^\alpha$, $\mmu$ have no first-principles microscopic definition, and are subject to redefinition ambiguities. These ambiguities are the reason why Eckart's formulation of dissipative relativistic hydrodynamics \cite{PhysRev.58.919} differs from the formulation of Landau and Lifshitz~\cite{LL6}. In MHD, similar redefinition ambiguities will affect the variables $\muphi$ and $h^\mu$ which encode the dynamics of the magnetic field~\cite{Grozdanov:2016tdf}. The idea of BDNK hydrodynamics~\cite{Kovtun:2019hdm, Bemfica:2019knx} is to use these redefinition ambiguities to address the causality problems of the original hydrodynamic theories \cite{PhysRev.58.919, LL6}. In practice, the procedure amounts to writing down all allowed one-derivative terms in the constitutive relations of conserved currents. The number of such terms will be greater than the number of physical transport coefficients; the latter are given by linear combinations of one-derivative parameters~\cite{Kovtun:2019hdm}. One then chooses those one-derivative terms which do not affect the physical transport coefficients in such a way that the resulting hydrodynamic equations are causal, and thermal equilibrium is stable. 

In general, the BDNK procedure is not guaranteed to work: there is no proof that for a relativistic dissipative fluid whose equation of state is thermodynamically stable, one can always write down causal and dynamically stable dissipative hydrodynamic equations by using the BDNK procedure.
If the BDNK construction happens to be successful (as it is for a number of physically relevant examples), then one has a viable dissipative hydrodynamic theory whose only equations of motion are conservation laws. Such equations can then be solved numerically, see e.g.~\cite{Pandya:2021ief, Pandya:2022pif, Bantilan:2022ech, Pandya:2022sff, Bea:2023rru}. 

The application of the above BDNK procedure to dMHD was discussed in reference~\cite{Armas:2022wvb} which studied dMHD without a neutral-particle current, and worked out conditions for equilibrium stability and linearized causality based on the first two equations in \eqref{eq:lin_caus_conds}. Here we will discuss dMHD with a neutral-particle current.
We will see that the last equation in \eqref{eq:lin_caus_conds}, a necessary condition for causality in linearized theories, is satisfied as well when the BDNK procedure is applied to dMHD.  We will then study the full non-linear causality of the resulting dMHD equations, and will show that the linearized analysis is in fact sufficient to demonstrate that the non-linear dMHD theory is causal. 

In order to write down the constitutive relations in dMHD, we need a list of all one-derivative scalars, vectors, and tensors made out of our chosen hydrodynamic variables $U$. We use the freedom of redefining $h^\mu$ out of equilibrium and demand that $u^\mu h_\mu = 0$ continue to hold in dMHD, like it does in perfect-fluid MHD. This choice simplifies the equations while still allowing for causal dynamics. With $u{\cdot}h=0$, one can use the decompositions \eqref{eq:TJJJ-general-uh} to write down the constitutive relations. Namely, the scalars ${\cal E}$, ${\cal P}$, ${\cal S}$, ${\cal Q}_\parallel$, ${\cal B}_\parallel$, ${\cal N}$, ${\cal J}_\parallel$ will be given by combinations of one-derivative scalars made out of $\nabla U$, the vectors ${\cal Q}^\mu_\perp$, ${\cal T}^\mu_\perp$, ${\cal B}^\mu_\perp$, ${\cal D}^\mu_\perp$, ${\cal J}^\mu_\perp$ will be given by combinations of one-derivative transverse vectors made out of $\nabla U$, and the tensors ${\cal T}^{\mu\nu}_\perp$, ${\cal D}^{\mu\nu}_\perp$ will be given by combinations of one-derivative transverse tensors made out of $\nabla U$. For fluids without a conserved neutral-particle current, the relevant one-derivative quantities were enumerated in reference~\cite{Hernandez:2017mch, Armas:2022wvb}.

\subsection{Dissipative dMHD}
\label{sec:diss-dmhd}
The procedure for writing down the one-derivative constitutive relations for $T^{\mu\nu}$, $J^{\mu\nu}$, and $J^\mu$ can be schematically described in two steps. First, one writes down the one-derivative hydrostatic free energy for fluid subject to static external sources: the metric $g_{\mu\nu}$ (coupled to $T^{\mu\nu}$), gauge field $\A_\mu$ (coupled to $J^\mu$), and antisymmetric $\Pi_{\mu\nu}$ (coupled to $J^{\mu\nu}$). Varying the hydrostatic free energy with respect to external sources gives rise to hydrostatic contributions in the constitutive relations, as was originally described in \cite{Jensen:2013vta, Banerjee:2012iz}. The second step is to write down all possible (allowed by the symmetry) non-equilibrium terms in the constitutive relations, with arbitrary coefficients. The physical transport coefficients can be constrained by demanding the positivity of entropy production, and can be related to response functions of the microscopic theory through Kubo formulas~\cite{Kovtun:2012rj}. For the current $J^\mu$ representing electrically neutral particles, the external source $\A_\mu$ is not physically realized, and should be set to zero for phenomenological applications. The external source $\Pi_{\mu\nu}$ can be related to external electric current, which is proportional to $\epsilon^{\mu\nu\rho\sigma}\nabla_{\!\nu} \Pi_{\rho\sigma}$. We will set $\Pi_{\mu\nu}$ to zero as well, when discussing dMHD equations. 

The non-equilibrium scalars, vectors, and tensors are obtained by projecting $\partial_{\mu}\alpha$,
$(\nabla_{\!\mu}\beta_\nu + \nabla_{\!\nu}\beta_\mu)$, $(\nabla_{\!\mu}\gamma_\nu - \nabla_{\!\nu}\gamma_\mu)$, where $\alpha \equiv \mmu/T$, $\beta^\mu \equiv u^\mu/T$, $\gamma^\mu \equiv H^\mu/T$, with $H^\mu \equiv \muphi h^\mu$. In equilibrium, the condition $\nabla_{\!\mu}\beta_\nu + \nabla_{\!\nu}\beta_\mu = 0$ (Killing's equation) is the covariant version of the statement that thermal equilibrium has constant temperature and velocity. Namely, the equilibrium fluid velocity is aligned with the timelike Killing vector, and the equilibrium temperature is given by the Tolman law~\cite{Jensen:2013vta}. Similarly, in equilibrium without external ``electric'' fields ${\mathtt E}_\mu$, the condition $\partial_{\lambda}\alpha = 0$ is simply the statement that both $T$ and $\mmu$ are proportional to $1/\sqrt{-g_{00}}$ \cite{LL5}, hence $\mmu/T$ is constant. Finally, in equilibrium without external $\Pi_{\mu\nu}$ fields, the condition $\nabla_{\!\mu}\gamma_\nu - \nabla_{\!\nu}\gamma_\mu = 0$ is the covariant version of the magnetostatic equation ${\bm\nabla} {\times} {\bf H} = 0$. See reference~\cite{Armas:2018zbe} for a discussion of hydrostatics of different phases of matter with external $\Pi_{\mu\nu}$ fields. The covariant version of the familiar magnetostatic equation ${\bm \nabla}{\cdot}{\bf B} = 0$ in curved space is $\nabla_{\!\alpha} B^\alpha  = B^\alpha \dot u_\alpha$, where $\dot u^\mu \equiv u^\lambda \nabla_{\!\lambda} u^\mu$ is the acceleration.%
\footnote{
  The full relation (which also holds out of equilibrium) is $\nabla_{\!\alpha} B^\alpha  = B^\alpha \dot u_\alpha - E_\alpha \Omega^\alpha$, where $\Omega^\mu = \epsilon^{\mu\nu\rho\sigma} u_\nu \nabla_{\!\rho} u_\sigma$ is the vorticity vector. In MHD, when the electric fields are not explicitly included, the term $E_\alpha \Omega^\alpha$ is dropped.
}

We will impose charge conjugation symmetry C on both hydrostatics and hydrodynamics. Physically, this corresponds to working with fluid states which are electrically neutral. The transformation is such that $h^\mu$ is C-odd, while $\muphi$ and $\rhophi$ are C-even (as can be read off from their expressions in terms of the magnetic field in perfect fluids). We will also assume parity-invariant microscopic dynamics. With these symmetries, the only one-derivative terms in the constitutive relations are non-hydrostatic~\cite{Grozdanov:2016tdf, Hernandez:2017mch, Armas:2018atq, Armas:2022wvb}. The most general one-derivative constitutive relations can then be written down as follows (again, assuming $u_\mu h^\mu = 0$). The parity-even scalars are:
\begin{align}
     s_1 = \frac{\dot T}{T} \,, \ \ \ \ 
     s_2 = \nabla_{\!\mu} u^\mu \,, \ \ \ \ 
     s_3 = u^\lambda \partial_{\lambda} \! \left( \frac{\mmu}{T} \right) , \ \ \ \ 
     s_4 = h^\mu h^\nu \nabla_{\!\mu} u_\nu \,,\ \ \ \
     s_5 =  \frac{T u^\lambda}{\muphi} \partial_{\lambda} \! \left( \frac{\muphi}{T} \right) ,
\end{align}
where $\dot T \equiv u^\lambda \partial_{\lambda} T$. 
The parity-odd scalars (pseudoscalars) are:
\begin{align}
     p_1 = h_\mu \left( \frac{\Delta^{\mu\lambda}\partial_{\lambda} T}{T} + \dot u^\mu \right) \,,\ \ \ \ 
     p_2 = \frac{1}{T \rhophi}\nabla_{\!\mu} (T \rhophi h^\mu)  \,, \ \ \ \ 
     p_3 = h^\lambda \partial_{\lambda} \! \left( \frac{\mmu}{T} \right) \,.
\end{align}
The vanishing of $p_2$ in equilibrium may be viewed as the equilibrium condition $\nabla_{\!\mu}(T B^\mu) = 0$.
The transverse parity-odd vectors are:
\begin{align}
   Y_1^\mu = T\Delta^{\mu\lambda}_\perp \left(  \partial_\lambda \! \left( \frac{\muphi}{T} \right) - \frac{\muphi}{T} h^\alpha \nabla_{\!\alpha} h_\lambda \right) ,\ \ \ \ 
   Y_2^\mu = T\Delta^{\mu\alpha}_\perp \left( \frac{\Delta^{\lambda}_{\alpha} \partial_{\lambda} T}{T} + \dot u_\alpha \right) ,\ \ \ \ 
   Y_3^\mu = T\Delta^{\mu\lambda}_\perp \partial_{\lambda} \!\left( \frac{\mmu}{T} \right) .
\end{align}
The transverse parity-even vectors (pseudo-vectors) are: 
\begin{align}
   \Sigma_1^\mu = \Delta^{\mu\alpha}_\perp h^\nu \left( \nabla_{\!\alpha} u_\nu + \nabla_{\!\nu} u_\alpha \right) , \ \ \ \ 
   \Sigma_2^\mu = \Delta^{\mu\alpha}_\perp u^\nu \left( \nabla_{\!\alpha} h_\nu - \nabla_{\!\nu} h_\alpha \right) .
\end{align}
There is one symmetric transverse traceless tensor: 
\begin{align}
  \sigma^{\mu\nu}_\perp = \left( \Delta^{\mu\alpha}_\perp \Delta^{\nu\beta}_\perp + \Delta^{\nu\alpha}_\perp \Delta^{\mu\beta}_\perp - \Delta^{\mu\nu}_\perp \Delta^{\alpha\beta}_\perp \right) \nabla_{\!\alpha} u_\beta  \,,
\end{align}
and one anti-symmetric transverse tensor:
\begin{align}
  Z^{\mu\nu} = \muphi \Delta^{\mu\rho}_\perp \Delta^{\nu\sigma}_\perp \left( \nabla_{\!\rho} h_\sigma - \nabla_{\!\sigma} h_\rho \right) .
\end{align}

We are now ready to write down the constitutive relations of dissipative MHD. Given the above non-equilibrium scalars, vectors, and tensors, the most general one-derivative constitutive relations for $T^{\mu\nu}$, $J^{\mu\nu}$, and $J^\mu$ are given by equations~\eqref{eq:TJJJ-general-uh} with the following constituents. The scalars are:%
\begin{subequations}
\label{eq:1-deriv-scalars}
\begin{align}
  & {\cal E} = \epsilon + \sum_{n=1}^5 \varepsilon_n s_n \,, \ \ \ \ 
    {\cal P} = p - \coeff13 \muphi \rhophi  + \sum_{n=1}^5 \pi_n s_n \,, \ \ \ \ 
    {\cal S} = -\coeff23 \muphi \rhophi + \sum_{n=1}^5 \sigma_n s_n \,,\\
  & {\cal B}_\parallel = \rhophi + \sum_{n=1}^5 \beta_{{\scriptscriptstyle\parallel} n} s_n \,,\ \ \ \ 
    {\cal N} = n + \sum_{n=1}^5 \nu_n s_n \,, \\
  & {\cal Q}_\parallel = \sum_{n=1}^3 \theta_{{\scriptscriptstyle\parallel} n} p_n \,,\ \ \ \ 
    {\cal J}_\parallel = \sum_{n=1}^3 \gamma_{{\scriptscriptstyle\parallel} n} p_n \,.
\end{align}
\end{subequations}
The vectors are:%
\begin{subequations}
\label{eq:1-deriv-vectors}
\begin{align}
  & {\cal Q}^\mu_\perp = \sum_{n=1}^3 \theta_{{\scriptscriptstyle\perp} n} Y^\mu_n \,,\ \ \ \ 
    {\cal D}^\mu_\perp = \sum_{n=1}^3 \rho_{{\scriptscriptstyle\perp} n} Y^\mu_n  \,,\ \ \ \ 
    {\cal J}^\mu_\perp = \sum_{n=1}^3 \gamma_{{\scriptscriptstyle\perp} n} Y^\mu_n \,, \\
  & {\cal T}^\mu_\perp = \sum_{n=1}^2 \tau_{n} \Sigma^\mu_n \,,\ \ \ \ 
    {\cal B}^\mu_\perp = \sum_{n=1}^2 \beta_{{\scriptscriptstyle\perp} n} \Sigma^\mu_n \,.
\end{align}
\end{subequations}
The tensors are:
\begin{align}
\label{eq:1-deriv-tensors}
  & {\cal T}^{\mu\nu}_\perp = -\eta_\perp \sigma^{\mu\nu}_\perp \,,\ \ \ \ 
    {\cal D}^{\mu\nu}_\perp = - r_\parallel Z^{\mu\nu} \,.
\end{align} 
The zero-derivative thermodynamic coefficients $\epsilon$, $p$, $\rhophi$, $n$ are functions of $(T, \mmu, \muphi)$, given by the equation of state. Alternatively, the equation of state may be expressed in terms of other variables, as was done for example in the ideal-gas expressions \eqref{eq:pfMHD-dual-3}. The one-derivative parameters $\varepsilon_n$, $\pi_n$, $\sigma_n$, $\beta_{{\scriptscriptstyle\parallel} n}$, $\nu_n$, $\theta_{{\scriptscriptstyle\parallel} n}$, $\gamma_{{\scriptscriptstyle\parallel} n}$, $\theta_{{\scriptscriptstyle\perp} n}$, $\rho_{{\scriptscriptstyle\perp} n}$, $\gamma_{{\scriptscriptstyle\perp} n}$, $\tau_{n}$, $\beta_{{\scriptscriptstyle\perp} n}$, $\eta_\perp$, $r_\parallel$ are functions of $(T, \mmu, \muphi)$, a total of 46 functions in dMHD; we shall call these ``transport parameters'', to distinguish them from physical ``transport coefficients'' such as resistivity.%
\footnote{
  The transport coefficients can be defined (and in principle, calculated) in the fundamental microscopic theory through correlation functions of quantum operators such as the energy-momentum tensor. The transport parameters, on the other hand, are merely auxiliary parameters in the effective hydrodynamic description of matter. Some hydrodynamic parameters, such as $\eta_\perp$, may happen to coincide with physical transport coefficients, but in general, transport parameters have no definition in the fundamental microscopic theory. 
}
The number of actual physical transport coefficients is much smaller.

\subsection{Frame invariants and transport coefficients}
Of the transport parameters given in \eqref{eq:1-deriv-scalars} - \eqref{eq:1-deriv-tensors}, only certain linear combinations are invariant under changes of hydrodynamic frame. Let us perform a first-order redefinition of the hydrodynamic variables, such that
\begin{equation}\label{eq:frame-transformation}
    \begin{split}
        T = T' + \delta T, \qquad \muphi &= \muphi' + \delta \muphi, \qquad \mmu = \mmu' + \delta \mmu\,,\\
        u^\mu = u'^\mu + \delta u_{\pparallel} h'^\mu + \delta u^\mu_{\pperp}, \quad & \quad h^\mu = h'^\mu + \delta u_{\pparallel} u'^\mu + \delta h^\mu_{\pperp}\,,
    \end{split}
\end{equation}
where the redefinition is such that $u^2=-1$, $h^2 = 1$, and $u{\cdot}h=0$ are respected up to second order in derivatives. Under such a redefinition, the transport parameters must change in order to keep the conserved one-point functions frame-invariant up to second order. The first-order redefinitions may in general be decomposed with respect to the basis quantities used to construct the first-order constitutive relations, i.e.
\begin{equation}
    \begin{split}
        \delta T = \sum_{i=1}^5 \delta \mathfrak{a}_i\,s_i, \qquad \delta \muphi &= \sum_{i=1}^5 \delta \mathfrak{b}_i\,s_i, \qquad \delta \mmu = \sum_{i=1}^5 \delta \mathfrak{c}_i\,s_i\,,\\
        \delta u_{\pparallel} = \sum_{j=1}^3 \delta \mathfrak{d}_j\,p_j, \qquad \delta u_{\pperp}^\mu &= \sum_{n=1}^3 \delta \mathfrak{e}_n\,Y_n^\mu, \quad \delta h_{\pperp}^\mu = \sum_{m=1}^2 \delta \mathfrak{f}_m\,\Sigma_m^\mu
    \end{split}
\end{equation}
 Defining $\delta \chi_i \equiv \chi'_i - \chi_i$ for transport parameter $\chi_i$, the scalar transport parameters\footnote{Here, ``scalar transport parameter" means ``transport parameter that multiplies a scalar first-order quantity in the constitutive relations."} transform according to
\begin{subequations}
    \begin{align}\label{eq:scalar-frame-transformation}
     \delta \varepsilon_i & =  \lr{\pder{\epsilon}{T} \delta \mathfrak{a}_i + \pder{\epsilon}{\muphi} \delta \mathfrak{b}_i + \pder{\epsilon}{\mmu} \delta \mathfrak{c}_i}\,,\\
        \delta\lr{\pi_i {-} \frac{1}{2} \sigma_i} & = \lr{\pder{p}{T} \delta \mathfrak{a}_i + \pder{p}{\muphi} \delta \mathfrak{b}_i + \pder{p}{\mmu} \delta \mathfrak{c}_i}\,,\\
        \delta\lr{\pi_i {+} \sigma_i} & = \biggl[\lr{\pder{p}{T} {-} \muphi \pder{\rhophi}{T}} \!\delta \mathfrak{a}_i + \lr{\pder{p}{\muphi} {-} \muphi \pder{\rhophi}{\muphi} {-} \rhophi} \!\delta \mathfrak{b}_i + \lr{\pder{p}{\mmu} {-} \muphi \pder{\rhophi}{\mmu}} \!\delta \mathfrak{c}_i\biggr],\\
       \delta\beta_{\pparallel i} & =  \lr{\pder{\rhophi}{T} \delta \mathfrak{a}_i + \pder{\rhophi}{\muphi} \delta \mathfrak{b}_i + \pder{\rhophi}{\mmu} \delta \mathfrak{c}_i}\,,\\
        \delta \nu_i & = \lr{\pder{n}{T} \delta \mathfrak{a}_i + \pder{n}{\muphi} \delta \mathfrak{b}_i + \pder{n}{\mmu} \mathfrak{c}_i}\,,
    \end{align}
\end{subequations}
where $i\in\{1..5\}$. The pseudoscalar transport parameters transform according to
\begin{equation}\label{eq:pseudoscalar-frame-transformation}
        \delta \theta_{\pparallel j} = \lr{\epsilon + p + \muphi \rhophi} \delta \mathfrak{d}_j, \qquad \delta \gamma_{\pparallel j} = n \,\delta \mathfrak{d}_j\,,
\end{equation}
where $j\in\{1..3\}$. The vector transport parameters transform according to
\begin{equation}
    \delta \theta_{\pperp n}  = \lr{\epsilon + p} \delta \mathfrak{e}_n,\qquad \delta \gamma_{\pperp n} = n \,\delta \mathfrak{e}_n, \qquad \delta \rho_{\pperp n} = \rho\, \delta \mathfrak{e}_n\,,
\end{equation}
where $n \in \{1..3\}$. The pseudovector transport parameters transform according to 
\begin{equation}
    \delta \tau_m = -\muphi \rhophi \,\delta \mathfrak{f}_m, \qquad \beta_{\pperp m} = \rhophi \,\delta \mathfrak{f}_m\,,
\end{equation}
where $m \in \{1..2\}$. The tensor transport parameters
\begin{equation}
    \eta_{\pperp}, \qquad r_{\pparallel}\,,
\end{equation}
are invariant under transformations of frame. It is possible to form frame-invariant linear combinations of the transport parameters~\cite{Kovtun:2019hdm, Hoult:2020eho, Armas:2022wvb}. There are 23 such frame invariants that may be constructed:
\begin{subequations}\label{eq:frame-invariants}
    \begin{align}
        f_i &\equiv \lr{\pi_i - \frac{1}{2} \sigma_i} - \left.\pder{p}{\epsilon} \right|_{(\rhophi,n)}  \,\ce_i - \left.\pder{p}{\rhophi}\right|_{(\epsilon,n)} \,\beta_{\pparallel i} - \left.\pder{p}{n}\right|_{(\epsilon,\rhophi)} \,\nu_i\,,\\
        g_i&\equiv \lr{\pi_i + \sigma_i} - \left.\pder{(p-\muphi \rhophi)}{\epsilon}\right|_{(\rhophi,n)} \,\ce_i - \left.\pder{(p-\muphi \rhophi)}{\rhophi}\right|_{(\epsilon,n)} \,\beta_{\pparallel i} \nonumber\\
        &- \,\left.\pder{(p-\muphi \rhophi)}{n}\right|_{(\epsilon,\rhophi)} \nu_i\,,\\
        h_j &\equiv \gamma_{\pparallel j} - \lr{\frac{n}{\epsilon + p - \muphi \rhophi}} \theta_{\pparallel j}\,,\\
        k_n &\equiv \rho_{\pperp n} - \lr{\frac{\rhophi}{\epsilon + p}}\theta_{\pperp n}\,,\\
        \ell_n &\equiv \gamma_{\pperp n} - \lr{\frac{n}{\epsilon + p}} \theta_{\pperp n}\,,\\
        m_m &\equiv \tau_m + \muphi \,\beta_{\pperp m}\,,\\
        &\,\,\eta_{\pperp},\qquad r_{\pparallel}\,.
    \end{align}
\end{subequations}
In dMHD without a neutral-particle current, there are 7 physical transport coefficients: two resistivities $r_\pperp$ and $r_\pparallel$ (transverse and longitudinal with respect to the magnetic field), the shear viscosity $\eta_\pperp$, and four other viscosities (rather, five viscosities, two of which are equal by the Onsager reciprocity relation)~\cite{Grozdanov:2016tdf, Hernandez:2017mch}. Even without taking into account the neutral-particle current, there are significantly more frame-invariant quantities than transport coefficients (14 of them~\cite{Armas:2022wvb}), and still more transport parameters.

The reason there are more transport parameters than transport coefficients has to do with the freedom of redefining hydrodynamic variables, and with the ``off-shell'' vs ``on-shell'' counting in the derivative expansion. See reference~\cite{Kovtun:2019hdm} for a discussion of how to express physical transport coefficients as linear combinations of transport parameters in normal fluids, and reference~\cite{Armas:2022wvb} for how to do so in dissipative dMHD. The idea of BDNK hydrodynamics is to use the freedom of choosing the transport parameters (in a way that does not affect the physical transport coefficients) so that the resulting hydrodynamic equations a) describe causal dynamics, and b) do not predict unphysical instabilities of the global equilibrium state. 

For dMHD with a neutral-particle current included, there are three scalars equations ($u_\nu \de_\mu T^{\mu\nu} = 0$, $h_\nu \de_\mu J^{\mu\nu} = 0$, and $\de_\mu J^\mu = 0$), two pseudoscalar equations ($h_\nu \de_\mu T^{\mu\nu} = 0$, $u_\nu \de_\mu J^{\mu\nu} = 0$), one vector equation ($\Delta^{\lambda}_{\perp,\nu} \de_\mu T^{\mu\nu} =0)$, and one pseudovector equation ($\Delta^{\lambda}_{\perp,\nu} \de_\mu J^{\mu\nu} = 0$). Going ``on-shell" with respect to these equations at ideal order, the basis scalars, pseudoscalars, vector, and pseudovectors become related to one another; therefore, the frame-invariants \eqref{eq:frame-invariants} become related to one another. We are then left with four scalar frame invariants (two each of the $f_i$, $g_i$), one pseudoscalar, four vectors (two each of the $\ell_n$, $k_n$), one pseudovector, and the two tensor frame invariants. This set is further related by two Onsager relations, leaving 10 physical transport coefficients, which may in principle be obtained from the underlying microscopic description through Kubo formulas of the linear response theory.

We define the analogue of the ``Landau frame'' convention \cite{LL6} for dMHD by demanding that the coefficients in the decomposition \eqref{eq:TJJJ-general-uh} satisfy
\begin{equation}
    {\cal E} = \epsilon, \quad {\cal B}_{\pparallel} = \rhophi, \quad {\cal N} = n, \quad {\cal Q}_{\pparallel} = 0, \quad {\cal Q}^\mu_{\pperp}  = 0, \quad {\cal B}^\mu_{\pperp} = 0\,.
\end{equation}
In this convention, the physical transport coefficients appear in the one-derivative constitutive relations \eqref{eq:dMHD-1} as
\begin{subequations}\label{eq:physical_transport_coefficients}
    \begin{align}
        T^{\mu\nu}_{(1)} &= -\lr{\zeta_{\pparallel} s_4 + \zeta_{\times} \lr{s_2 {-} s_4}} h^\mu h^\nu - \lr{\zeta_{\times} s_4 + \zeta_{\pperp} \lr{s_2 {-} s_4}} \Delta^{\mu\nu}_{\pperp} - \eta_{\pparallel} \Sigma_1^{(\mu} h^{\nu)} - \eta_{\pperp} \sigma^{\mu\nu}_{\pperp} ,\\
        J^{\mu\nu}_{(1)} &= - r_{\pperp} Y_1^{[\mu} h^{\nu]} - \tilde{\sigma} Y_3^{[\mu} h^{\nu]} - r_{\pparallel} Z^{\mu\nu} \,,\\
        J^\mu_{(1)} &= - \lr{\sigma_{\pparallel} p_3} h^\mu - \tilde{\sigma} Y_1^\mu - \sigma_{\pperp} Y_3^\mu \,.
    \end{align}
\end{subequations}
We note the physical transport coefficient $\tilde{\sigma}$, which was absent in previous formulations of dMHD with a U(1) current. This coefficient describes the response of the neutral-particle flux to a transverse gradient of $B^2$. The fact that the same $\tilde \sigma$ appears in the constitutive relations for both $J^{\mu\nu}$ and $J^\mu$ is a consequence of the Onsager relations. The expressions for the physical transport coefficients in \eqref{eq:physical_transport_coefficients} in terms of the frame invariants \eqref{eq:frame-invariants} may be found by applying the ideal-order equations in the Landau frame to eliminate $s_1, s_3, s_5, p_1, p_2, Y_2^\mu$, and $\Sigma_2^\mu$. A few relatively simple examples are given by $r_{\pperp}$, $\eta_{\pparallel}$, and $\sigma_{\pperp}$:
\begin{equation}
\label{eq:example_physical}
        r_{\pperp} = - \lr{k_1 - \lr{\frac{\rhophi T}{\epsilon + p}} k_2}, \qquad \eta_{\pparallel} = - \lr{ m_1 - m_2} , \qquad \sigma_{\pperp} = - \lr{\ell_3 - \frac{n T}{\epsilon + p} \ell_2}\,.
\end{equation}
The physical transport coefficients are subject to the inequality-type constraints  
\begin{equation}
    \begin{gathered}
        \eta_{\pparallel} \geq 0 \qquad \eta_{\pperp} \geq 0, \qquad \sigma_{\pparallel} \geq 0, \qquad r_{\pparallel} \geq 0\,,\\
        \zeta_{\pparallel} \geq 0, \qquad \zeta_{\pperp} \geq 0, \qquad \zeta_{\pparallel} \zeta_{\pperp} \geq \zeta_{\times}^2\,,\\
        \sigma_{\pperp} \geq 0, \qquad r_{\pperp} \geq 0, \qquad \sigma_{\pperp} r_{\pperp} \geq \tilde{\sigma}^2\,,
    \end{gathered}
\end{equation}
due to demanding the non-negativity of entropy production.

In summary, the equations of dissipative dMHD are the conservation laws \eqref{eq:TJJJ-conserv} together with the constitutive relations \eqref{eq:TJJJ-general-uh}, \eqref{eq:1-deriv-scalars}, \eqref{eq:1-deriv-vectors}, \eqref{eq:1-deriv-tensors}. The dynamical variables are $T$, $\muphi$, $\mmu$, $u^\lambda$, and $h^\lambda$, which satisfy $u^2=-1$, $h^2=1$, $u{\cdot}h = 0$. The equations have 46 transport parameters, however only 10 combinations of the transport parameters are physical transport coefficients ($\eta_\pperp$, $\eta_\pparallel$, $\zeta_\pperp$, $\zeta_\pparallel$, $\zeta_\times$, $r_\pperp$, $r_\pparallel$, $\sigma_\pperp$, $\sigma_\pparallel$, $\tilde\sigma$). The rest of the transport parameters are not physical transport coefficients, but rather serve as ``small-scale regulators'', to be chosen so that the resulting differential equations respect causality, and the equilibrium state is stable. We now turn to causality properties of the dissipative dMHD equations. 

\section{Causality in dissipative dMHD}
\label{sec:equiv_dMHD}
\subsection{Linearized analysis}
\label{sec:lin_analysis}
Let us begin by considering linearized plane-wave perturbations about a homogenous equilibrium state. Perturbations about the fluid rest frame are described by
\begin{equation}
    \begin{gathered}
    T = T_0 + \delta T, \ \ \ \  \muphi = \muphi_0 + \delta \muphi\,, \ \ \ \  \mmu = \mmu_0 + \delta \mmu\,, \\
    u^\mu = \delta^\mu_t + \delta u_{\pparallel} \delta^\mu_z + \delta u_{\pperp}^\mu\,\ \ \ \ 
    h^\mu = \delta^\mu_z + \delta u_{\pparallel}  \delta^\mu_t + \delta h^\mu_{\pperp} \,,
    \end{gathered}
\end{equation}
where the background values $T_0$, $\muphi_0$, $\mmu_0$ are constant, and the perturbations are proportional to $\exp(-i\omega t + i k_i x^i)$. In the fluid rest frame, we have chosen the $z$-axis along $h^\mu$, the transverse perturbations $\delta u^\mu_{\pperp}$, $\delta h^\mu_{\pperp}$ both lie in the $xy$-plane. One can use the residual $SO(2)$ symmetry to orient $k_j$ in the $xz$-plane, so that $k_i x^i =  k \sin(\theta) x + k \cos(\theta) z$, where $k \equiv |{\bf k}|$,
and $\theta$ is the angle between the wave vector ${\bf k}$ and the background magnetic field.
Upon inserting these perturbations into the equations of motion \eqref{eq:TJJJ-conserv} with the constitutive relations \eqref{eq:TJJJ-general-uh}, \eqref{eq:1-deriv-scalars} -- \eqref{eq:1-deriv-tensors} and linearizing, we obtain the spectral curve $F(\omega,k)$ as the determinant of the coefficient matrix of $\delta U^B = \{\delta T, \delta \muphi, \delta \mmu, \delta u_{\pparallel}, \delta u_{\pperp}^\mu, \delta h_{\pperp}^\mu\}$. The eigenmodes of the linearized dMHD fluctuations are determined by $F(\omega, k) = 0$.

Due to the theory respecting $P$ and $C$ symmetries, as well as the SO(2) symmetry of the equilibrium state, the spectral curve factorizes into an Alfv\'en channel and a diffusion-magnetosonic channel:
\begin{equation}
    F(\omega,k) = F_{\rm Alfv\acute{e}n}(\omega,k) F_\textrm{d-ms}(\omega,k)\,.
\end{equation}
The spectral curve $F_{\rm Alfv\acute{e}n}(\omega,k)$ is a 4-th order polynomial in $\omega$, whose coefficients depend on the transport parameters $\{\theta_{\pperp 1,2}, \rho_{\pperp 1,2}, \tau_{1,2}, \beta_{\pperp 1,2}, \eta_{\pperp}, r_{\pparallel}\}$. The spectral curve $F_\textrm{d-ms}(\omega,k)$ is a 12-th order polynomial in $\omega$, whose coefficients depend on all of the transport parameters except for $r_{\pparallel}$. 
We will briefly discuss stability and causality in each channel. We leave a more thorough (possibly numerical) investigation to future work.

\subsubsection{Alfv\'en channel}
The Alfv\'en channel describes the mixing of the $y$-component of the magnetic field, and the $y$-component of the fluid velocity. The spectral curve is a quartic polynomial in $\omega$, whose coefficients are functions of $k$, $\theta$, and the background parameters. It satisfies the third causality criterion of \eqref{eq:lin_caus_conds}. 

In the limit $k \to 0$, there are two gapless (hydrodynamic) modes, and two gapped (non-hydrodynamic) modes. At small $k$, they behave as%
\footnote{
   The dispersion relations $\omega(k)$ exhibit a non-commutativity of limits $k\to 0$ and $\theta\to\pi/2$ \cite{Hernandez:2017mch,Fang:2024hxa,Fang:2024skm}. The same non-commutativity does not arise for $k\to\infty$ and $\theta\to\pi/2$ at leading order in large~$k$.
}
\begin{subequations}
    \begin{align}
\label{eq:Alfven-wave-dispersion}
        \omega(k) &= \pm \sqrt{\frac{\muphi \rhophi}{p+\epsilon}} \cos(\theta) k - \frac{i}{2} \lr{\Gamma_{A1} \sin(\theta)^2 + \Gamma_{A2} \cos(\theta)^2} k^2 + O(k^3) \,, \\
        \omega(k) &= - i \lr{\frac{p + \epsilon}{\theta_{\pperp 2}}} + O(k^2) \,, \\
        \omega(k) &= i \lr{\frac{\rhophi}{\beta_{\pperp 2}}} + O(k^2) \,,
    \end{align}
\end{subequations}
where
\begin{equation}
\label{eq:Gamma-Alfven}
        \Gamma_{A1} =  \frac{\eta_{\perp}}{p+\epsilon} + \frac{r_{\pparallel} \muphi}{\rhophi} \,,\ \ \ \ 
        \Gamma_{A2} = \frac{\eta_{\pparallel}}{p+\epsilon} + \frac{r_{\pperp} \muphi}{\rhophi} \,,
\end{equation}
and we have written the transport parameters that appear in the hydrodynamic modes in terms of the physical transport coefficients. In the limit of large~$k$, we find that all four of the modes have linear dispersion relations, $\omega(k) \sim c \,k$. The values of $c$ are given by the roots of a quadratic equation in $c^2$:
\begin{equation}
\label{eq:Alfven_poly}
    P_{\rm Alfv\acute{e}n}(c^2) \equiv \lr{- \beta_{\pperp 2} \theta_{\pperp 2}} c^4 + a c^2 + b = 0 \,,
\end{equation}
where
\begin{equation}
    \begin{split}
        a &= \frac{1}{T} \lr{\beta_{\pperp 1} \theta_{\pperp 1} \muphi + \theta_{\pperp 2} \rho_{\pperp 1} \muphi - \theta_{\pperp 1} \rho_{\pperp 2} \muphi - T \beta_{\pperp 2} \tau_1 + T \lr{\beta_{\pperp 1} - \rho_{\pperp2}} \tau_2} \cos(\theta)^2 \\
        &+ \lr{\beta_{\pperp 2} \eta_{\perp} - r_{\parallel} \theta_{\pperp 2} \muphi} \sin(\theta)^2\,,\\
        b &= \frac{\muphi}{4 T} \lr{ - r_{\pparallel} T + \rho_{\pperp 1} + \lr{r_{\pparallel} T + \rho_{\pperp 1}}\cos(2\theta)}\lr{-\eta_{\pperp} + \tau_1 + \lr{\eta_{\pperp} + \tau_1} \cos(2\theta)}\,.
    \end{split}
\end{equation}
Unlike the small-$k$ Alfv\'en wave dispersion relation \eqref{eq:Alfven-wave-dispersion}, the large-$k$ equation~\eqref{eq:Alfven_poly} can not be expressed solely in terms of the physical transport coefficients.
Demanding causality amounts to demanding that the roots of equation~\eqref{eq:Alfven_poly} be real, and $-1\leq c\leq 1$. In order to enforce these conditions, one may of course simply solve equation~\eqref{eq:Alfven_poly} directly and then impose constraints on the transport parameters which would ensure $-1\leq c\leq 1$. A more comprehensive method is as follows.

We may ensure that $0 \leq c^2 < 1$ by imposing various constraints on the coefficients of the polynomial~\eqref{eq:Alfven_poly}. In order to ensure that the roots are real, one may enforce positivity of the discriminant. Ensuring $c^2\geq 0$ may be enforced by means of the Routh-Hurwitz (RH) conditions~\cite{korn2013mathematical,Hastir_2023} applied to the transformed polynomial $P_{\rm Alfv\acute{e}n}(-c^2)$  (the RH criteria enforce that the roots have negative real part). Finally, one may enforce that $c^2 < 1$ by demanding Schur-Cohn stability \cite{Gargantini-Schur,Zahreddine1992}, which ensures that the roots of~\eqref{eq:Alfven_poly} lie within a disc of radius one of the origin in the plane of complex $c^2$. Schur-Cohn stability is achieved by enforcing the RH criteria on the M\"obius-transformed polynomial $P'(c^2) = (c^2-1)^2 P_{\rm Alfv\acute{e}n}\lr{(c^2+1)/(c^2-1)}$; the M\"obius transformation maps the unit disc to the left-hand complex half-plane. 
For the explicit RH conditions on polynomials of orders two, three, four, and six (with all non-zero coefficients), see appendix~B of~\cite{Hoult2020:thesis}.

As the inclusion of the neutral-particle current in dMHD does not modify the equations in the Alfv\'en channel, the linearized causality conditions found in~\cite{Armas:2022wvb} apply here. Translating these to our notation, it is sufficient to impose
\begin{equation}
\label{eq:Alfven-causality}
    \begin{gathered}
        \theta_{\pperp 1} = - \frac{\tau_2}{\muphi} \quad \text{or} \quad \beta_{\pperp 1} =  T \rho_{\pperp 2}\,,\\
        T \theta_{\pperp 2} \rho_{\pperp 1} - \frac{\tau_1}{\muphi} \beta_{\pperp 2} + \rho_{\pperp 1} \tau_1 > 0, \quad r_{\pparallel} \eta_{\pperp} - T r_{\pparallel} \theta_{\pperp 2} + \frac{\eta_{\pperp}}{\muphi} \beta_{\pperp 2} > 0\,,\\
        r_{\pparallel} \eta_{\pperp} + \eta_{\pperp} \rho_{\pperp 1} + r_{\pparallel} \tau_1 + \rho_{\pperp 1} \tau_1 < 0\,,\\
        \theta_{\pperp 2} > 0 \,, \qquad \frac{\beta_{\pperp 2}}{\rhophi} < 0\,,
    \end{gathered}
\end{equation}
where the final two conditions arise from demanding stability of the gapped modes. In the Landau frame formulation~\eqref{eq:physical_transport_coefficients}, one has $\beta_{\pperp 2} = \theta_{\pperp 2} = 0$. One subsequently finds that the causality conditions~\eqref{eq:lin_caus_conds} are violated in the Landau frame: at large~$|\bf{k}|$, there are only two modes $\omega = \omega(\bf{k})$, with $\omega$ proportional to $ i |\bf{k}|^2$.

\subsubsection{Diffusion-magnetosonic channel}
The $x$ and $z$ components of the magnetic field, the temperature, the $x$ and $z$ components of the fluid velocity, and the neutral particle chemical potential all mix together in the diffusion-magnetosonic channel. The spectral curve $F_\textrm{d-ms}(\omega,k)$ is a $12$-th order polynomial in $\omega$ which satisfies the third causality condition of \eqref{eq:lin_caus_conds}. 

In the limit of small-$k$, there are six non-hydrodynamic modes, and six hydrodynamic modes. Three of the non-hydrodynamic modes are straightforward to write down:
\begin{equation}
\label{eq:cms-gaps}
    \omega = - i \lr{\frac{p + \epsilon}{\theta_{\pperp 2}}} + O(k^2)\, , \quad \omega = i \lr{\frac{\rhophi}{\beta_{\pperp 2}}} + O(k^2)\, , \quad \omega = - i \lr{\frac{p + \epsilon - \muphi \rhophi}{\theta_{\pparallel 1} + \theta_{\pparallel 2}}} + O(k^2)\,.
\end{equation}
Two of the above zero-$k$ frequencies are the same as in the Alfv\'en channel.%
\footnote{
  The reason for this is that in the limit $\theta \to 0$ (or $\theta \to \pm \pi$) the SO($2$) symmetry is restored, and the spectral curve $F_\textrm{d-ms}$ factors into a second copy of $F_{\rm Alfv\acute{e}n}$ and a remainder. 
}
The other three non-hydrodynamic (gapped) frequencies are given by the roots of a cubic polynomial 
\begin{equation}\label{eq:P3g-explicit}
\begin{split}
    P_{3g}(\omega) &= \det\left| \begin{pmatrix}
        \pder{\epsilon}{T} & \pder{\epsilon}{\muphi}& \pder{\epsilon}{\mmu}\\
        \pder{\rhophi}{T} & \pder{\rhophi}{\muphi} & \pder{\rhophi}{\mmu}\\
        \pder{n}{T} & \pder{n}{\muphi} & \pder{n}{\mmu}
    \end{pmatrix} + \lr{- i \omega} \begin{pmatrix}
        \frac{\lr{\ce_1 - \ce_5}T-\ce_3 \mmu}{T^2} & \frac{\ce_5}{\mmu} & \frac{\ce_3}{T}\\
        \frac{\lr{\beta_{\pparallel 1} - \beta_{\pparallel 5}}T - \beta_{\pparallel 3} \mmu}{T^2} & \frac{\beta_{\pparallel 5}}{\mmu} & \frac{\beta_{\pparallel 3}}{T}\\
        \frac{\lr{\nu_1 - \nu_5}T - \nu_3 \mmu}{T^2} & \frac{\nu_5}{\mmu} & \frac{\nu_3}{T}
    \end{pmatrix}
    \right|\\
    &= a (-i \omega)^3 + b (-i \omega)^2 + c (-i \omega) + d\,,
\end{split}
\end{equation}
which defines $a$, $b$, $c$ as functions of the transport parameters and thermodynamic quantities. In order to ensure the stability of $P_{3g}(\omega)$, one can ensure the Routh-Hurwitz stability of $P_{3 g}(i \Delta)$ in $\Delta \equiv -i \omega$. The polynomial $P_{3 g}(i \Delta)$ is RH stable if
\begin{equation}
    a > 0 \qquad b > 0, \qquad d>0, \qquad b c - a d > 0\,,
\end{equation}
where the sign is fixed by the demand that the Jacobian $\partial(\epsilon, \rhophi, n)/\partial(T,\muphi,\mmu)$ is positive.

Of the six hydrodynamic modes, one is identically zero ($\omega = 0$ for all $k$). Four are the magnetosonic modes, given by
\begin{equation}
    \omega = \pm {\cal V}_{\pm} k - i \Gamma_{\pm} k^2 + O(k^3)\,,
\end{equation}
where ``$+$'' in ${\cal V}_{\pm}$, $\Gamma_{\pm}$ corresponds to the ``fast" magnetosonic modes, and ``$-$'' corresponds to the ``slow" magnetosonic modes. 
The final hydrodynamic mode is the diffusion mode, with
\begin{equation}
    \omega = - i \Gamma_q k^2 + O(k^4)\,.
\end{equation}
The wave speeds ${\cal V}_{\pm}$ and the damping coefficients $\Gamma_\pm$, $\Gamma_q$ can all be expressed in terms of physical transport coefficients and thermodynamic quantities. 

In the limit of large-$k$, all modes have linear dispersion relations, $\omega \sim c k$. The asymptotic phase velocities $c$ are given by the roots of an order-6 polynomial in $c^2$. The polynomial factors, such that the phase velocities are determined by
\begin{equation}\label{eq:d-ms-factor}
   c^2 P_\textrm{d-ms}(c^2) = 0 ,
\end{equation}
where $P_\textrm{d-ms}(c^2)$ is an order-5 polynomial in $c^2$. It is possible to choose a frame in which this quintic further factorizes -- this ``decoupled frame", in the spirit of the decoupled frame in \cite{Hoult:2020eho}, separates the diffusive mode out from the magnetosonic mode at large $k$. The decoupled frame is found by
\begin{equation}
    \ce_3 = 0, \quad \theta_{\pparallel 2} = 0, \quad\theta_{\pparallel 3} = 0, \quad \theta_{\pperp 3} = 0, \quad \pi_3 = 0, \quad \sigma_3 = 0, \quad \rho_{\pperp 3} = 0, \quad \beta_{\pparallel 3} = 0\,.
\end{equation}
In this frame, $P_\textrm{d-ms}$ factorizes into
\begin{equation}
    P_\textrm{d-ms}(c^2) = \lr{c^2 \nu_3 \rhophi + \lr{\pder{\rho}{\mmu} T \gamma_{\pparallel 2} + \gamma_{\pparallel 3} \rhophi} \cos^2(\theta) + \gamma_{\pperp3} \rhophi \sin^2(\theta)} P_{\rm ms}(c^2) \,,
\end{equation}
where $P_{\rm ms}(c^2)$ is a quartic polynomial in $c^2$. The asymptotic phase velocities $c$ are determined by solving $ P_\textrm{d-ms}(c^2) = 0$; the causality requirement that $0 \leq c^2 < 1$ for all $\theta$ then imposes constraints on the transport parameters. 
The roots of $P_{\rm ms}(c^2)$ can be rendered causal in a similar fashion to the Alfv\'en channel; by demanding Schur-Cohn stability of the roots, demanding the Routh stability of $P_{\rm ms}(-c^2)$ , and demanding reality of the roots (which for a quartic equation requires slightly more work than just positivity of the discriminant).

Finally, two of the modes in $F_\textrm{d-ms}$ are non-propagating at large $k$ (as described by the overall factor of $c^2$ in equation \eqref{eq:d-ms-factor}). These two modes correspond to the overall zero mode ($\omega = 0$ for all $k$), and one of the gapped modes which solves $P_{3g}(\omega)=0$. These two modes are fictitious, in the sense that they are a product of not taking the constraint equation into account when doing the analysis. If the constraint equation is included in the linearized analysis, these two modes disappear from the spectrum. To make this more apparent, one can show that there exists a frame (the ``constraint frame") in which the two non-propagating modes appear as decoupled factors in the spectral curve for all (not just large)~$k$.%
\footnote{
As an example, choosing to write the solution to the ideal dMHD constraint equation in the rest frame as 
\[
\begin{gathered}
\delta \muphi = \pder{\rhophi}{T} \delta \phi(\omega,\k), \qquad  \delta \mmu = \pder{\rhophi}{T} \delta \psi(\omega,\k),\qquad \delta h^x(\omega,\k) = \pder{\rhophi}{T} \cos(\theta) \delta \chi(\omega,\k)\,,\\
\delta T(\omega,\k) = - \pder{\rhophi}{\muphi} \delta\phi(\omega,\k) - \pder{\rhophi}{\mmu} \delta \psi(\omega,\k) - \rhophi \sin(\theta) \delta \chi(\omega,\k)\,,
\end{gathered}
\]
where $\delta \phi(\omega,k)$, $\delta \psi(\omega,k)$, $\delta \chi(\omega,k)$ are a new set of degrees of freedom, the transport parameters in the constraint frame are given by
\[
\begin{gathered}
    \beta_{\pparallel 1} = -\frac{{\cal D}_{\rhophi}}{\rhophi} \beta_{\pparallel 4}, \quad \beta_{\pparallel 2} = - \beta_{\pparallel 4}, \quad \beta_{\pparallel 3} = - \frac{T}{\rhophi} \pder{\rhophi}{\mmu} \beta_{\pparallel 4}, \quad \beta_{\pparallel 5} = - \frac{\muphi}{\rhophi} \pder{\rhophi}{\muphi} \beta_{\pparallel 4}\,,\\
    \beta_{\pperp 1} = \beta_{\pparallel 4}, \quad \beta_{\pperp 2} = \beta_{\pparallel 4}\,,
\end{gathered}
\]
where ${\cal D}_{\rhophi} \equiv \pder{\rhophi}{T} T + \pder{\rhophi}{\muphi} \muphi + \pder{\rhophi}{\mmu} \mmu$. 
}
In this frame, the constraint equation is solved (in the linearized theory, in the fluid rest frame) by the same relation between variables which solves the constraint (in the linearized theory, in the fluid rest frame) in ideal dMHD. If one includes the constraint, these factorized modes vanish from the spectrum (so that $F_\textrm{d-ms}$ becomes an order-10 polynomial in $\omega$).
We note that the decoupled frame and the constraint frame are in general not consistent with each other unless the transport parameter $\beta_{\pparallel 4}$ vanishes.

\subsection{Non-linear causality}
\label{sec:nlc}
We now turn to analyzing the causality of the non-linear system of partial differential equations for dissipative dMHD, equations~\eqref{eq:TJJJ-conserv} with the constitutive relations~\eqref{eq:TJJJ-general-uh}, \eqref{eq:1-deriv-scalars} - \eqref{eq:1-deriv-tensors}. We will use the results of section~\ref{sec:caus_and_eq} to show that if the linearized theory is found to be causal, the non-linear theory is also causal.

Let us assume that after performing the linearized analysis of section~\ref{sec:lin_analysis}, the linearized theory of perturbations about equilibrium has been made causal via an appropriate choice of the transport parameters. 
We then note that the equations of motion are of the form~\eqref{eq:quasilin_gen}. In particular, in $D$ spacetime dimensions, there is one constraint equation, and $2D$ dynamical equations. Decomposing the equations of motion into dynamical and constraint equations, one finds that the equations of dissipative dMHD are of the form~\eqref{eq:dyn_con_sufficient}, with $N_{\rm c}=1$. In other words, the sufficient conditions a) and b) of section~\ref{sec:caus_and_eq} are both satisfied for dissipative dMHD.

The final question with regards to the applicability of the linear vs non-linear causality equivalence is the constraint equation. Since the sufficient conditions a) and b) hold, the applicability of the equivalence may be determined by inspecting the large-$k$ limit of the spectral curve $F(\omega,\bf{k})$ in the linearized analysis. One can work with coordinates in which $n^\mu = (1,{\bf 0})$, take $\omega \to \lambda \omega$, $k\to\lambda k$, and expand the spectral curve at large $\lambda$. The highest-order (in $\lambda$) term in the spectral curve has one overall factor of $\omega$ at non-zero fluid velocity ${\bf v}$. The rest of the highest-order (in $\lambda$) term is a Lorentz scalar. Therefore, by direct computation, we find that the characteristic equation is of the form~\eqref{eq:good_form_characteristic}, with $\ell = 1$.

Due to the fact that the sufficient conditions a) and b) of section~\ref{sec:caus_and_eq} hold for dMHD, and the characteristic equation is of the form~\eqref{eq:good_form_characteristic}, we conclude that if the linearized theory is made causal, the non-linear equations of motion for dMHD with a neutral-particle current will also be causal.

Interestingly, despite the fact $\ell=1$, there are two non-propagating (in the sense that $\Re(\omega)/|{\bf k}| \to 0$ at large $k$) modes that appear when linearizing about the fluid rest frame (${\bf v}{=}0$). This corresponds to an overall factor of $(n{\cdot}\xi) (u{\cdot}\xi)$ in the characteristic equation.  When one solves the constraint equation (in the fluid rest frame, using the ``constraint frame") in the linearized analysis, both non-propagating modes disappear from the spectrum; yet, only one of the two modes is non-covariant. The interplay between constraint equations and non-propagating modes is something to which we plan to return in later work.

\section{Conclusions}\label{sec:conclusion}
Let us summarize our results. We have discussed how for systems of second-order quasilinear equations, one may relate their causality properties to the causality properties of linearized equations which describe small fluctuations about a homogeneous solution. The linearized system will correctly reflect the causality properties of the full non-linear system provided {\em a)} the principal part of the non-linear system does not depend on derivatives of the fields, and {\em b)} the homogeneous solution about which the linearization occurs is not subject to algebraic constraints, except for those that also apply to the fields themselves. 

We then investigated relativistic dissipative magnetohydrodynamics supplemented by a conserved current of electrically neutral particles (sometimes called a ``mass current'' in the literature). We adopted the formulation of dissipative magnetohydrodynamics in which all equations take the form of exact conservation laws, a formulation which we called ``dMHD". With the neutral-particle current taken into account, dMHD has 10 physical transport coefficients which include anisotropic viscosities, anisotropic electric resistivities, and anisotropic particle-number conductivities. Using a convention for hydrodynamic variables which is analogous to the Landau-Lifshitz convention in normal-fluid hydrodynamics~\cite{LL6}, the constitutive relations with the 10 transport coefficients are given by equation~\eqref{eq:physical_transport_coefficients}.

Even though the analogue of the Landau-Lifshitz convention allows for a direct enumeration of transport coefficients in equation~\eqref{eq:physical_transport_coefficients}, the convention gives rise to acausal hydrodynamic equations. We thus looked at an application of the so-called BDNK recipe \cite{Bemfica:2017wps, Kovtun:2019hdm, Bemfica:2019knx, Hoult:2020eho} to dissipative dMHD, a procedure which may render first-order hydrodynamic theories%
\footnote{
   First-order in the sense that the constitutive relations for conserved currents contain only terms with up to one derivative of the hydrodynamic variables. The hydrodynamic equations (conservation laws) in first-order theories are thus second-order partial differential equations. 
}
such as dissipative dMHD causal and stable.
We found that dissipative dMHD satisfies the sufficient conditions {\em a)} and {\em b)} above; therefore, a linearized analysis of perturbations about equilibrium is sufficient to ensure causality of the full non-linear theory. The equations of dissipative dMHD are second-order partial differential equations with derivative constraints; we found that the presence of constraints is not an obstacle to the linearized analysis reproducing the correct causality conditions. 

There are several open avenues of investigation. In our analysis, we only considered systems of second-order differential equations. While a generalization to systems of higher-order equations is in principle straightforward, it would be interesting to study mixed-order systems, in which different degrees of freedom obey differential equations of differing order. 
Regarding the equivalence between causality of linearized equations, and causality of the full quasilinear equations, we only considered characteristic equations of the form \eqref{eq:good_form_characteristic}. We leave a more general investigation to future work, including the interplay of constrained dissipative fluid-dynamical systems with gravity. 

Another direction which we plan to investigate in the future is the actual stability and causality conditions of dissipative dMHD with a neutral-particle current; in this paper, we simply noted the equivalence that exists between the linearized and non-linear versions of the theory. Given a particular equation of state, the linearized analysis can be performed numerically in order to determine classes of hydrodynamic frames which are stable and causal. Ensuring stability for an anisotropic state is in general less straightforward than in the isotropic case. For an isotropic equilibrium state, the spectral curve $F(\omega,\k)$ is a polynomial in $(-i\omega)$ with real coefficients that are functions of $k^2$. In the anisotropic case, the spectral curve is a polynomial in $(-i \omega)$ whose coefficients are functions of $k^2$ and $i{\bf k}{\cdot}{\bf B}$. As the coefficients of the polynomial in $(-i\omega)$ are in principle complex, the standard Routh-Hurwitz criterion would no longer apply, and one would have to use one of its complex extensions. Here, we assumed parity invariance, hence the coefficients of the polynomial are real functions of $k^2$ and $({\bf k}{\cdot}{\bf B})^2$, and the standard Routh-Hurwitz criterion still applies. 

Finally, there remains the problem of ensuring local well-posedness of the system of partial differential equations. 
The first step is to determine whether the system of equations is strongly hyperbolic. 
We leave the analysis of strong hyperbolicity of dissipative dMHD and its causal extensions to future work.

\paragraph{Acknowledgements:} This work was supported in part by an Alexander Graham Bell Doctoral Scholarship from the Natural Sciences and Engineering Research Council of Canada (NSERC). This work was also supported in part by the National Science Foundation under Grant No. NSF PHY-1748958. REH would like to thank J. Noronha for hosting him at the University of Illinois at Urbana-Champaign, where a portion of the work for this paper was completed. PK would like to thank the Dutch Institute for Emergent Phenomena (Amsterdam) for hospitality. The authors would like to thank D.~Almaalol, J.~Armas, M.~Disconzi, L.~Gavassino, and J.~Noronha for helpful conversations.

\bibliographystyle{JHEP}
\bibliography{hydro-general-biblio}

\end{document}